\documentclass{article}
\usepackage{jheppub}

\usepackage{bm}
\usepackage{amsmath}
\usepackage{mathtools}
\usepackage{float}
\usepackage[english]{babel}
\usepackage{caption}
\usepackage{subcaption}

\def\qq{\tilde q_0}
\def\qqq{\tilde q_1}
\newcommand{\m}{{\bar{m}}}
\newcommand{\mm}{{\bar{m}_1}}

\def\q{\bar{q}}
\def\b{\beta^{'}}
\newcommand{\SOMMA}[2]{\displaystyle\sum\limits_{#1}^{#2}}

\newcommand{\bsigma}{{\boldsymbol \sigma}}

\newcommand{\si}{\sigma_i}

\newcommand{\resub}{\textcolor{black}}

\newcommand{\bra}{{\langle}}
\newcommand{\ket}{{\rangle}}

\newcommand{\sech}{\  \textnormal{sech}}

\title{About the de Almeida-Thouless line in neural networks}
\author[a,b,d,e]{L. Albanese} 
\author[c,e]{A. Alessandrelli}
\author[b]{A. Annibale}
\author[a,e] {A. Barra}

\affiliation[a] {Dipartimento di Matematica e Fisica ``Ennio De Giorgi'', Universit\`a del Salento, Lecce, Italy.}
\affiliation[b]  {Department of Mathematics, King's College London, Strand, London WC2R 2LS, UK.} 
\affiliation[c]{Dipartimento di Informatica, Università di Pisa, Lungarno Antonio Pacinotti, 43, 56126, Pisa, Italy.}
\affiliation[d] {Scuola Superiore ISUFI, Universit\`a del Salento, Lecce, Italy. }
\affiliation[e] {Istituto Nazionale di Fisica Nucleare, Sezione di Lecce, Italy.}

\abstract{
{In this work we present a rigorous and straightforward method to detect the onset of the instability of replica-symmetric theories in information processing systems, which does not require 
a full replica analysis 
as in the method originally proposed by de Almeida and Thouless for spin glasses. The method is based on an expansion of the free-energy obtained within one-step of replica symmetry breaking (RSB) around the RS value. As such, it requires solely continuity and differentiability of the free-energy and it is robust to be applied broadly to systems with quenched disorder. 
We apply the method to the Hopfield model and to neural networks with multi-node Hebbian interactions, as case studies. In the appendices we test the method on the Sherrington-Kirkpatrick and the Ising $P$-spin models, recovering the AT lines known in the literature for these models, as a special limit, which corresponds to assuming that the transition from the RS to the RSB phase can be obtained by varying continuously the order parameters. 
Our method provides a generalization of the AT approach, which does not rely on this limit and can be applied to systems with discontinuous phase transitions, as we show explicitly for the spherical P-spin model, recovering the known RS instability line. %Our results suggest that the AT line does not accurately capture the RS instability of the Hopfield model in the retrieval phase.
}
}

\begin{document}

\maketitle

\noindent
\section{Introduction}
% \old{More and more papers in recent Literature are paying attention to replica symmetry breaking in neural networks (see e.g. \cite{Albanese2021,Zecchina-Pnas2016,Zecchina-PRL2021,Haiping-PRR2023,Gavin-PRE2022}), yet -at present- neither a general broken replica-symmetry theory for these systems is available nor a systematic method to find the emergence of replica-symmetry instability.} $\to$
{Replica symmetry breaking in neural networks has attracted  increasing attention in recent years  \cite{Albanese2021,Zecchina-Pnas2016,Zecchina-PRL2021,Haiping-PRR2023,Gavin-PRE2022}, however there is, as yet, no general broken replica-symmetry theory for these systems and  
no simple method to systematically detect their transition from the replica-symmetric (RS) to the replica-symmetry-broken (RSB) phase.}

% \old{While the first point is still out of reach, as for the latter we can instead adapt the classical approach developed for standard spin glasses: indeed after the seminal work by de Almeida and Thouless in 1978 \cite{de1978stability} on the stability of the replica-symmetric picture for the Sherrington-Kirkpatrick spin glass model, many researchers have continued to inspect the emergence of replica symmetry breaking as a departure from replica symmetry making their techniques more and more rigorous along these decades (see e.g. \cite{talagrand2003spin, guerra2006replica, chen2021almeida,Holler-PRE2020,Chokri-JSP2022,Charbonneau-PRE2019,Kondor-arxiv2022}) and testing various models (see e.g.).} $\to$

While the first point is still out of reach, the second question can be addressed by adapting approaches originally developed for spin glasses. Indeed, the instability line of the RS phase in the Sherrington-Kirkpatrick (SK) spin glass model was derived by de Almeida and Thouless (AT) many decades ago \cite{de1978stability}, using a method based on replicas. Since their seminal work, rigorous techniques have been developed and tested in archetypical mean-field as well as short-ranged 
spin glass models, by many researchers (see e.g. \cite{talagrand2003spin, bardina2004, guerra2006replica, chen2021almeida,Holler-PRE2020,Chokri-JSP2022,Charbonneau-PRE2019,Kondor-arxiv2022}).

As neural networks are particular realizations of spin glasses, it is quite natural to ask if we can devise a systematic method to derive the RS instability line %(alternative to the heuristic original one by de Almeida $\&$ Thouless) 
also for these systems.
In this work we answer affirmatively to this question, using the Hopfield model of neural networks and a model of dense associative memory, which extends Hebbian learning to multi-node interactions, as case studies. To this purpose, we devise a method inspired by the approach proposed by Toninelli in \cite{toninelli2002almeida}, which builds on Guerra's 
work on broken replica-symmetry bounds \cite{guerra_broken}. 
As a technical note we remark that at difference with conventional spin glasses, here we focus on the RS instability in the parameter space $(\alpha, T)$ where $\alpha$ is the storage load of the network and $T$ is the noise level, rather than in the space $(h,T)$ (i.e. magnetic field, temperature) conventionally used in spin glasses. 

%For the Hopfield model, \linda{in a particular limit}, we recover the same instability line obtained by Coolen \cite{coolen2001statistical} using the AT approach. \linda{Therefore, our method gives an expression which is more general than the aforementioned one.}

For the Hopfield model, our method recovers the instability line obtained by Coolen \cite{coolen2001statistical} using the AT approach, as a special limit, which corresponds to assuming a continuous transition from the RS to the RSB phase, in the order parameters. Therefore, our method provides a generalization of the AT approach, 
which can be applied to systems 
with a discontinuous transition from the RS 
to the 1RSB phase. 
Another advantage of our method, when compared to the involved calculations of the AT method in the Hopfield model \cite{coolen2001statistical}, is its remarkable simplicity. This allows for straightforward application to more complex neural network models, such as dense associative memories with $P$-node interactions. 
We supplement the results in the main text with appendix \ref{app:spin-glasses} where we show our method at work on conventional spin-glass models, namely the Sherrington-Kirkpatrick model, the Ising $P$-spin and the spherical $P$-spin model, the latter providing an  example of system which exhibits a discontinuous phase transition from the RS to the RSB phase. In all cases, we retrieve the AT lines known for these models \cite{de1978stability, gardner1985spin, Crisanti, coolen2001statistical} in a specific limit, confirming the validity of our approach as a generalization of the AT method. As expected from the decomposition theorem of multi-node Hebbian networks proved in \cite{Albanese2021}, for dense associative memories with $P$-node interactions we retrieve the instability line of the Ising $P$-spin model derived in Appendix \ref{app:spin-glasses}. 
Appendices \ref{app:expt0} and \ref{app:subleading} provide further technical details.

\section{The Hopfield model}
\label{sec:Hopfield}
\par\medskip
% \old{Let us start describing the first model: the Hopfield network is the theoretical prototype for pattern recognition by neural networks endowed with associative memory. By associative memory we mean the ability of the network to reconstruct names, objects, faces, scheme, etc. (i.e. \textit{patterns of information} generally speaking) starting from incomplete or corrupted data supply \cite{Amit,Coolen}. 
% %
% In order to describe analytically this network, we consider } $\to$
{In this section we illustrate the method for the 
Hopfield model with $N$ Ising neurons $\si \in \{1,-1 \}, \ i=1, \hdots, N$ and $K=\alpha N$ stored patterns $\bm \xi^\mu$, $\mu=1,\ldots, K$. Each pattern $\bm\xi^\mu$ is a sequence of $N$ Rademacher entries (i.e. Bernoulli variables) $\xi_i^\mu$, $i=1, \hdots, N$, with distribution}
\begin{align}
    \mathbb{P}(\xi_i^\mu)=\dfrac{1}{2}\left( \delta_{\xi_i^\mu, +1} + \delta_{\xi_i^\mu, -1}\right).
\end{align}
The Hamiltonian of the model is 
\begin{align}
    H_N(\bm \sigma \vert \bm \xi)=-\dfrac{1}{N} \sum_{i,j =1, 1}^{N, N} \sum_{\mu=1}^K \xi^\mu_i \xi^\mu_j \si \sigma_j
    \label{eq:HamiltonianHOP}
\end{align}
and we denote the associated Boltzmann factor, at inverse temperature $\beta=1/T$, as
\begin{equation}
\mathcal{B}_N(\bm{\sigma}|\bm{\xi})=\frac{e^{-\beta H_N(\bm \sigma \vert \bm \xi)}}{Z},\quad\quad\quad Z=\sum_{\bsigma}e^{-\beta H_N(\bm \sigma \vert \bm \xi)}.
\end{equation}

% \old{Moreover, it is convenient to introduce the order parameters $m$ and $q_{12}$ as
% \begin{eqnarray}
% \label{eq:def_m}
%          m&:=\dfrac{1}{N}\SOMMA{i=1}{N}\xi_i^{1}\sigma_i \ \ \ \ \ \ 
%          q_{12}&:=\dfrac{1}{N}\SOMMA{i=1}{N}\sigma_i^{(1)}\sigma_i^{(2)}, 
% \end{eqnarray}
% where the first one, denoted as the \textit{Mattis magnetization}, quantifies the alignment of the network configuration with the pattern $\boldsymbol \xi^1$ (i.e. its retrieval) and $q_{12}$ is the standard \textit{two-replica overlap} between two different replicas, namely two neural configurations of the network obtained with the same  coupling's realization $J_{ij}=\sum_{\mu=1}^K \xi^\mu_i \xi^\mu_j$.}
% $\to$
{In the so-called `retrieval' phase, the equilibrium local configurations are correlated only with a 
single pattern, say $\nu$. As the couplings $J_{ij}=\sum_{\mu=1}^K \xi^\mu_i \xi^\mu_j$ are symmetric w.r.t. permutations of the patterns, it is assumed without loss of generality that 
$\nu=1$. It is then convenient to define 
\iffalse
the following order parameters:
\begin{eqnarray}
\label{eq:def_m}
         m&\!\!:=\dfrac{1}{N}\SOMMA{i=1}{N}\xi_i^{1} \sigma_i \quad\quad 
         q&\!\!:=\dfrac{1}{N}\SOMMA{i=1}{N}         \sigma_i^{(1)}\sigma_i^{(2)}
\end{eqnarray}
where the first one, denoted as the \textit{Mattis magnetization}, quantifies the alignment of the 
%equilibrium
network configuration with the retrieval pattern $\boldsymbol \xi^1$ and $q$ is the standard \textit{two-replica overlap} quantifying the correlations between two 
%equilibrium
configurations of the network obtained with the same coupling's realization. 
Denoting their equilibrium values with
\begin{eqnarray}
         \m&\!\!:=\dfrac{1}{N}\SOMMA{i=1}{N}\xi_i^{1} \bra \sigma_i \ket\quad\quad 
         \q&\!\!:=\dfrac{1}{N}\SOMMA{i=1}{N}         \bra \sigma_i^{(1)}\ket \bra \sigma_i^{(2)}\ket 
         \nonumber
\end{eqnarray}
where $\bra \cdot \ket$ is the average with respect to the Boltzmann distribution with Hamiltonian (\ref{eq:HamiltonianHOP}) and noise level $T=1/\beta$, 
\fi
as the order parameters of the system, the so-called 
{\it Mattis magnetization}
\begin{eqnarray}
\label{eq:def_m}
m(\bsigma):=\dfrac{1}{N}\SOMMA{i=1}{N}\xi_i^{1} \sigma_i 
\end{eqnarray}
which quantifies the alignment of the 
system configuration $\bsigma$ with the retrieval pattern $\boldsymbol \xi^1$, and the \textit{two-replica overlap} 
\begin{eqnarray}
q(\bsigma^{(1)},\bsigma^{(2)})\!\!:=\dfrac{1}{N}\SOMMA{i=1}{N}\sigma_i^{(1)}\sigma_i^{(2)}
\end{eqnarray}
which quantifies the correlations between two 
configurations $\bsigma^{(1)},\bsigma^{(2)}$ of the system, with the same realization of the patterns (i.e. quenched disorder).

The RS analysis assumes that the order parameters 
$m$ and $q$ self-average around their equilibrium values $\m$ and $\q$, in the thermodynamic limit, namely

\begin{align} \label{eq:P_RS_Hop1}
    \lim_{N \rightarrow + \infty} P_N(m) =& \delta (m - \bar{m}),  \\ \label{eq:P_RS_Hop2}
    \lim_{N \rightarrow + \infty} P'_N(q) =& \delta (q - \bar{q}),
\end{align}
\resub{where $P_N(m)=\mathbb{E}_{\bm{\xi}}\sum_{\bm{\sigma}}\mathcal{B}_N(\bm{\sigma}|\bm{\xi})
\delta (m-m(\bm{\sigma}))
$ and 
$P_N'(q)=\mathbb{E}_{\bm{\xi}}\sum_{\bsigma^{(1)}, \bsigma^{(2)}}\mathcal{B}_N(\sigma^{(1)}|\bm{\xi})\mathcal{B}_N(\bsigma^{(2)}|\bm{\xi})
\delta (q-q_{12}(\bsigma^{(1)},\bsigma^{(2)}))
$, with
$\mathbb{E}_{\bm{\xi}}$ denoting the 
expectation over the pattern distribution (or 'quenched' disorder). }
%
% \old{and the expression of the related quenched statistical pressure of the model reads as} $\to$
Under this assumption, the free-energy, averaged over the pattern distribution, $f$, is given by (see \cite{Amit}) 
\begin{align}
    %\mathcal{A}_{RS}(\beta, \alpha \vert \q) 
    -\beta f_{RS}(\m, \bar{q} \vert \beta, \alpha)
    =& \ln 2 - \dfrac{\alpha}{2} \ln (1-\beta(1-\q)) - \dfrac{\beta}{2}\m^2 + \dfrac{\alpha \beta \q}{2(1-\beta(1-\q))} \notag \\
    &- \dfrac{\alpha\beta^2}{2} \dfrac{\q(1-\q)}{(1-\beta(1-\q))^2} + \mathbb{E} \ln \cosh \left( \beta z \sqrt{\dfrac{\alpha \q}{(1-\beta(1-\q))^2}} + \beta \m\right),
\end{align}
where $z$ is a random Gaussian variable with 
zero average and unit variance, $\mathbb{E}$ denotes the average over $z$ and $\alpha$ is the load capacity of the network. In this limit, the order parameters $\q$ and $\m$ fulfill the celebrated 
% \old{AGS} 
Amit-Gutfreund-Sompolinsky self-consistency equations \cite{Amit,Coolen}: 
\begin{align} %\label{eq:self_RS_Hop}
    \q&= \mathbb{E} \tanh^2 \left( \beta \m + \beta \sqrt{\dfrac{\alpha \q}{(1-\beta(1-\q))^2}} z\right),
    \label{eq:AGSq}\\
      \m &= \mathbb{E} \tanh \left( \beta \m + \beta \sqrt{\dfrac{\alpha \q}{(1-\beta(1-\q))^2}} z\right). \label{eq:AGSm} %\notag 
\end{align}   
% \old{The standard scenario with one-step of replica symmetry breaking (1-RSB) \cite{Crisanti-RSB,Steffan-RSB,AABO-JPA2020} assumes instead that, while $m$ still self-averages at $\m$,} $to$
On the other hand, within one step of the replica-symmetry breaking (1RSB) scheme 
\cite{Crisanti-RSB,Steffan-RSB,AABO-JPA2020}
it is assumed that
  the distribution of the two-replica overlap $q$, in the thermodynamic limit, displays two delta-peaks at the equilibrium values $\q_0$ and $\q_1>\q_0$ and the concentration on these two values is ruled by the parameter $\theta \in [0, 1]$, while $m$ self-averages as in the RS case : 
\begin{align}
    \resub{\lim_{N \rightarrow + \infty} P_N(m)} &\resub{= \delta (m - \bar{m}_1)}, \label{eq:1RSBAssm}  \\
    \lim_{N \rightarrow + \infty} P'_N(q) &= \theta \delta (q - \bar{q}_0) + (1-\theta) \delta (q - \bar{q}_1),     \label{eq:1RSBAss} 
\end{align}  
% \old{As we need it for comparison with the RS-counterpart, for the sake of brevity  we directly write down also the 1-RSB approximation of the quenched statistical pressure  related to the model \eqref{eq:HamiltonianHOP}: this reads as} $\to$
Within this assumption, the disorder-averaged free-energy is given by (see e.g. \cite{AABO-JPA2020})
%\footnote{In order to recover this expression, one can use any suitable technique of disordered statistical mechanics, e.g. Guerra's interpolation as in \cite{AABO-JPA2020}.}
\begin{align}
\label{eq:A1RSBHOP}
    %&\mathcal{A}_{1RSB}(\beta, \alpha, \theta \vert \q_0, \q_1) 
    -\beta f_{1RSB}(\mm, \bar{q}_1,\bar{q}_0 \vert \beta, \alpha, \theta)
    &=\ln 2 - \dfrac{\alpha}{2} \ln \left( \Delta_1 (\beta, \q_1)\right) + \dfrac{\alpha}{2\theta} \ln \left(\dfrac{\Delta_1 (\beta, \q_1) }{ \Delta_2(\beta, \theta,\q_0, \q_1)}  \right)+ \dfrac{\alpha \beta \q_0}{2\Delta_2(\beta, \theta, \q_0, \q_1)} 
    \notag \\
    & - \dfrac{\beta}{2} \mm^2+\frac{1}{2} \alpha \beta^2 \theta \frac{\q_0^2}{\Delta_2^2(\beta, \theta,\q_0, \q_1)} +\dfrac{1}{\theta} \mathbb{E}_1 \ln \mathbb{E}_2 \cosh^\theta g_\theta(\beta, \alpha, \mm, \q_0, \q_1) \notag \\
    &-\frac{1}{2} \alpha \beta^2 \theta \q_1 \!\left(\frac{\q_0}{\Delta_2^2(\beta, \theta, \q_0, \q_1)}+\frac{\q_1-\q_0}{\Delta_1 (\beta, \q_1)  \Delta_2(\beta, \theta, \q_0, \q_1)}\!\right)\notag \\
        &-\frac{1}{2} \alpha \beta^2 (1\!-\!\q_1) \!\left(\frac{\q_0}{\Delta_2^2(\beta, \theta, \q_0, \q_1)}\!+\!\frac{\q_1-\q_0}{\Delta_1 (\beta,\q_1) \Delta_2(\beta, \theta, \q_0, \q_1)}\!\right)
\end{align}
\normalsize
% \end{strip}
where, for mathematical convenience, we defined 
\begin{align}
&\Delta_1 (\beta, \q_1):= 1-\beta (1-\q_1)
\label{eq:DElta1}
\\
&\Delta_2(\beta, \theta,\q_0, \q_1):= 1-\beta (1-\q_1)-\beta \theta (\q_1-\q_0)
\label{eq:Delta2}\\
%\end{align}
%and
%\begin{align}
&g_\theta(\beta, \alpha, \mm, \q_1, \q_0):= \beta \mm+\frac{\beta z^{(1)}\sqrt{\alpha \q_0}}{\Delta_2(\beta, \theta, \q_0, \q_1)} + \beta z^{(2)} \sqrt{ \frac{\alpha(\q_1-\q_0)}{\Delta_1(\beta, \q_1)  \Delta_2(\beta, \theta,\q_0, \q_1)}} 
\label{eq:gtheta}
\end{align}
and we have denoted with $\mathbb{E}_1$, $\mathbb{E}_2$ the averages w.r.t. the standard normal variables $z^{(1)}$ and $z^{(2)}$, respectively. From now on, we imply the dependence of all the functions on $ \beta$ and $\alpha$.
By extremizing the 1RSB 
%statistical pressure 
free-energy
w.r.t. its order parameters  $\q_0, \ \q_1, \ \mm$, it is possible to show that the latter fulfill the following self-consistency equations 
\begin{equation}
\begin{array}{lll}
    \mm &= \mathbb{E}_1 \left[ \dfrac{\mathbb{E}_2 \cosh^\theta {g_\theta(\resub{\mm,\,} \q_1, \q_0)} \tanh {g_\theta(\resub{\mm,\,} \q_1, \q_0)}}{\mathbb{E}_2 \cosh^\theta {g_\theta(\resub{\mm,\,} \q_1, \q_0)}}\right], \\\\
    \q_1 &= \mathbb{E}_1 \left[ \dfrac{\mathbb{E}_2 \cosh^\theta {g_\theta(\resub{\mm,\,} \q_1, \q_0)} \tanh^2 {g_\theta(\resub{\mm,\,} \q_1, \q_0)}}{\mathbb{E}_2 \cosh^\theta {g_\theta(\resub{\mm,\,} \q_1, \q_0)}}\right], \\\\
    \q_0 &= \mathbb{E}_1 \left[ \dfrac{\mathbb{E}_2 \cosh^\theta {g_\theta(\resub{\mm,\,} \q_1, \q_0)} \tanh {g_\theta(\resub{\mm,\,} \q_1, \q_0)}}{\mathbb{E}_2 \cosh^\theta {g_\theta(\resub{\mm,\,} \q_1, \q_0)}}\right]^2.
\end{array}
\label{eq:self_RSBHOP}
\end{equation}
The key idea of our method is to assume that at the onset of the RS instability, one of the two delta-peaks in equation \eqref{eq:1RSBAss}
has vanishing weight, i.e. either $\theta\!\to\! 0$ or $\theta\!\to\! 1$. Consistency with the RS theory then requires the dominating peak to be located at the value $\q$ of the RS order parameter, so either $\lim_{\theta\to 0}\q_1\!=\!\q$  or 
$\lim_{\theta\to 1}\q_0\!=\!\q$. As we will see below, both these 
relations are generally satisfied, hence we appeal to the physical interpretation of RS breaking to determine which scenario applies. 
Noting that when the RS theory becomes unstable, a multiplicity of states emerges with mutual overlap $\q_0$ and self-overlap $\q_1$, within
the single state (or each of the states) predicted by the RS 
theory, it is natural to assume that the value of the mutual overlap between the newly born states is 
equal to the self-overlap $\q$ of the state(s) assumed by the RS theory. Thus, we will 
assume that at the onset of the RS instability, $\q_0\to \q$, hence  
$\theta\to 1$.
For the Hopfield model, taking this limit in  \eqref{eq:self_RSBHOP}, and 
using  
\begin{align}
\label{eq:g1HOP}
g_1(\resub{\mm,\,} \q_1, \q_0) = \beta \mm + \beta z^{(1)} \dfrac{\sqrt{\alpha \q_0}}{\Delta_1(\q_0)} + \beta z^{(2)} \sqrt{\dfrac{\alpha(\q_1-\q_0)}{\Delta_1(\q_1)\Delta_1(\q_0)}},
\end{align}
we get 
\begin{align}
    \lim_{\theta\to 1}\q_0&=  
%\mathbb{E}_1 \left[ \dfrac{\mathbb{E}_2 \cosh {g_1(\resub{\mm,\,} \q_1, \q_0)} \tanh {g_1(\resub{\mm,\,} \q_1, \q_0)}}{\mathbb{E}_2 \cosh {g_1(\resub{\mm,\,} \q_1, \q_0)}}\right]^2= 
\mathbb{E}_1 \left[ \dfrac{\mathbb{E}_2 \sinh {g_1(\resub{\mm,\,} \q_1, \q_0)}}{\mathbb{E}_2 \cosh {g_1(\resub{\mm,\,} \q_1, \q_0)}}\right]^2 \nonumber \\
&= \mathbb{E}_1 \left[ \dfrac{ \exp \left( \dfrac{\beta^2 \alpha(\q_1-\q_0)}{2 \Delta_1(\q_1)  \Delta_1(\q_0)}\right)\sinh\left( \beta \mm + \beta \sqrt{\dfrac{\alpha \q_0}{(1-\beta(1-\q_0))^2}} z^{(1)}\right)}{\exp \left( \dfrac{\beta^2 \alpha(\q_1-\q_0)}{2 \Delta_1(\q_1)  \Delta_1(\q_0)}\right) \cosh \left( \beta \mm + \beta \sqrt{\dfrac{\alpha \q_0}{(1-\beta(1-\q_0))^2}} z^{(1)}\right)}\right]^2 \nonumber \\
    &=\mathbb{E}_1 \tanh^2 \left( \beta \mm + \beta \sqrt{\dfrac{\alpha \q_0}{(1-\beta(1-\q_0))^2}} z^{(1)}\right)%\equiv\bar{q}
    \label{eq:q-lim-theta1}
\end{align}
where in the second line we have used $\sinh(A+B)=\sinh A \cosh B+\sinh B \cosh A$ and $\cosh(A+B)=\cosh A \cosh B+\sinh B \sinh A$ ($A$ denoting the first two terms on the RHS of \eqref{eq:g1HOP} and $B$ the last one) and have performed the integral over $z^{(2)}$ using the oddity of the $\sinh$ function.
% used that for $\theta=0$, $\Delta_2(0, \q_0, \q_1)=\Delta_1(\q_1)$ and the relation 
% \begin{align}
% \label{eq:relation}
%     \mathbb{E}_{\lambda,Y}[F(a_1+ \lambda a_2+Y a_3)]=\mathbb{E}_{_Z}\left[F\left(a_1+Z\sqrt{a_2^2+a_3^2}\right)\right],
% \end{align}
% with $F$ any smooth function, $a_1, \ a_2,\ a_3 \in \mathbb{R}$, and $\lambda$, $Y$ and $Z$ i.i.d. standard normal random variables. 
As \eqref{eq:q-lim-theta1}
 is identical to \eqref{eq:AGSq}, in the limit $\theta\to 1$, $\q_0$ is indeed equal to the RS order parameter $\q$, as anticipated.
%, $\lambda$, $Y$ $Z \sim \mathcal{N}(0,1)$. 
\resub{Similarly, we can show that 
\begin{align}
    \lim_{\theta\to 1}\mm&=\m 
    \end{align}
}
and one can easily verify that $f_{1RSB}(\mm,\bar{q}_{1},\bar{q}_0|\theta)|_{\theta=1}=f_{RS}(\m,\q)$, as expected from the fact that, for $\theta=1$, Eq. (\ref{eq:1RSBAss}) reduces to (\ref{eq:P_RS_Hop2}) and one retrieves the RS scheme.
Our purpose is then to prove that for values of $\theta$ close but away from one, the 1RSB expression of the quenched 
%statistical pressure 
free-energy is 
%smaller 
smaller
than the RS expression, i.e. 
$f_{1RSB}(\mm,\bar{q}_{1},\bar{q}_0|\theta)<f_{RS}(\m,\bar{q})$, 
below a critical line in the parameters space $(\alpha,\beta)$.  

To this purpose, we expand the 1RSB quenched 
free-energy 
around $\theta=1$ (i.e. around the replica symmetric expression) to the first order, writing
\begin{align} 
   f_{1RSB} (\resub{\mm,\,} \bar{q}_1,\bar{q}_0 \vert \theta)=& f_{1RSB}(\resub{\mm,\,} \bar{q}_1,\bar{q}_0 \vert \theta)|_{\theta=1}
   + (\theta-1) \partial_\theta f_{1RSB}(\resub{\mm,\,} \bar{q}_1,\bar{q}_0 \vert \theta)\vert_{\theta=1},
    \label{eq:expansionHOP_t1}
\end{align}
where $f_{1RSB}(\resub{\mm,\,} \bar{q}_1,\bar{q}_0\vert \theta)|_{\theta=1}= f_{RS}(\resub{\m,\,} \bar{q})$. To determine when the RS solution becomes unstable, i.e.  $f_{1RSB}(\resub{\mm,\,} \bar{q}_1,\bar{q}_0\vert \theta)<f_{RS}(\resub{\m,\,} \bar{q})$ we inspect the sign of $\partial_\theta f_{1RSB}(\resub{\mm,\,} \bar{q}_1,\bar{q}_0\vert \theta)\vert_{\theta=1}$, keeping in mind that $\theta-1 <0$.
To evaluate the latter, we need to expand the self-consistency equations for \resub{$\mm,$} $\q_0$ and $\q_1$ around $\theta=1$, to linear orders in $\theta-1$. 
We obtain
\begin{align}
\label{eq:q0_t1}
    \q_0 &= \mathbb{E}_1 \tanh^2 \left( \beta \m + \beta \sqrt{\dfrac{\alpha \q_0}{(1-\beta(1-\q_0))^2}} z^{(1)}\right) 
    + (\theta-1) A(\resub{\mm,\,} \q_0,\q_1) 
%    \notag \\%+\mathcal{O}(\theta^2) =  
%    &=: \mathbb{E} \tanh^2 \left( \beta \m + \beta J \sqrt{\dfrac{\alpha \q_1}{1-\beta(1-\q_1)}}z\right) + \theta A(\q_1, \q_0) 
    \end{align}
    where $A(\resub{\mm,\,} \q_0,\q_1)$ is a function of $\resub{\mm}$, $\q_0$ and $\q_1$ that will drop out of the calculation, whose expression is provided in \eqref{eq:A-Hopfield_t1}.
%
%It follows from \eqref{eq:q-lim-theta1} that 
As to $\mathcal{O}((\theta-1)^0)$, $\q_0$ \resub{and $\mm$ are equal to the RS order parameters $\q$ and $\m$, respectively,}
    we can rewrite \eqref{eq:q0_t1} as 
    \begin{equation}
        \q_0= \q+(\theta-1) A(\resub{\m,\,} \q, \q_1).
        \label{eq:q1-q-pre_t1}
    \end{equation}
    Following the same path for $\q_1$, and using \eqref{eq:q1-q-pre_t1}, we have 
    % \begin{align}
    % \q_0 =& \mathbb{E}_1\left( \mathbb{E}_2 \tanh g(\q_0, \q_1) \right)^2 + 2 \theta \notag \\
    % &\cdot\Big\{ \mathbb{E}_1 \left[ \mathbb{E}_2 \tanh g(\q_0, \q_1) \mathbb{E}_2 \tanh g(\q_0, \q_1) \ln \cosh g(\q_0, \q_1)\right] \notag \\
    % &- \mathbb{E}_1 \left[ \left(\mathbb{E}_2 \tanh g(\q_0, \q_1)\right)^2 \mathbb{E}_2 \ln \cosh g(\q_0, \q_1) \right] \notag \\
    % &
    % %\textcolor{white}{\mathbb{E}_1\left( \mathbb{E}_2 \tanh g \right)^2 + 2 \theta}
    % + \dfrac{\beta^3 (\q_1-\q_0)\alpha \q_0 }{\Delta_1(\qq, \q)^3} \mathbb{E}_1 \left[ \mathbb{E}_2 (1-\tanh^2 g) \mathbb{E}_2 \tanh^2 g \right] \notag \\
    % &
    % %\textcolor{white}{\mathbb{E}_1\left( \mathbb{E}_2 \tanh g \right)^2 + 2 \theta}
    % + \dfrac{2\beta^3 \alpha \q_1 (\q_1\!-\!\q_0) }{\Delta_1(\q, \qq)^3} \mathbb{E}_1 \left[ \mathbb{E}_2 \tanh g \mathbb{E}_2 \tanh g (1\!-\!\tanh^2 g) \right]\!\!\Big\} \notag \\
    % =:& \mathbb{E}_1\left( \mathbb{E}_2 \tanh g(\q_0, \q_1) \right)^2 + \theta B(\q, \qq)
    % \end{align}
    \begin{align}
    \q_1
    %&= \mathbb{E}_1\left( \mathbb{E}_2 \tanh g(\q_0, \q) \right)^2 + \theta B(\q_1,\q_0)\notag \\
    &=: \mathbb{E}_1 \left[ \dfrac{\mathbb{E}_2 \cosh {g_1(\resub{\m,\,} \q_1, \q)} \tanh^2 {g_1(\resub{\m,\,} \q_1, \q)}}{\mathbb{E}_2 \cosh {g_1(\resub{\m,\,} \q_1, \q)}}\right] + (\theta-1) B(\resub{\mm,\,} \q_0, \q_1)
    \label{eq:expanded-self_t1}
    \end{align}
where $B(\resub{\mm,\,} \q_0,\q_1)$ is provided in \eqref{eq:B-Hopfield_t1} and will drop out of the calculation.
%    Denoted with $\qq$ the r.h.s. first term, we have that,
For $\theta=1$, we have
\begin{align}
    \q_1= \mathbb{E}_1 \left[ \dfrac{\mathbb{E}_2 \cosh {g_1(\resub{\m,\,} \q_1, \q)} \tanh^2 {g_1(\resub{\m,\,} \q_1, \q)}}{\mathbb{E}_2 \cosh {g_1(\resub{\m,\,} \q_1, \q)}}\right] %= \qq
    \label{eq:self2_t1}
\end{align}
which is a self-consistency equation for $\q_1$, that depends only on $\q$ \resub{and $\m$}. Denoting with $\tilde q_1(\resub{\m,\,} \q)$ its solution, 
we can then write \eqref{eq:expanded-self_t1} as 
\begin{align}
        \q_1= \qqq(\resub{\m,\,} \q)+(\theta-1) B(\resub{\m,\,} \q, \qqq(\resub{\m,\,} \q))
        \label{eq:q1-q_t1}
\end{align}
and, finally, 
    \begin{equation}
        \q_0= \q+(\theta-1) A(\resub{\m,\,} \q,\qqq(\resub{\m,\,} \q)).
        \label{eq:q0-q_t1}
    \end{equation}
\resub{Similarly to $\q_0$ and $\q_1$ we can expand also $\mm$ as}
\begin{align}
    \mm = \m + (\theta-1) C( \resub{\m,\,} \q,\qqq(\resub{\m,\,} \q))
    \label{eq:m_t1}
\end{align}
\resub{where $C( \resub{\m,\,} \q,\qqq(\resub{\m,\,} \q))$ is provided in \eqref{eq:C-Hopfield_t1}.}

Using \eqref{eq:m_t1}, \eqref{eq:q1-q_t1} and \eqref{eq:q0-q_t1} to  evaluate the derivative of 
$f_{1RSB}(\resub{\mm,\,} \bar{q}_1,\q_0 \vert \theta)$
w.r.t. $\theta$ and finally setting  $\theta=1$, we obtain:
\begin{align}
    K(\resub{\m,\,} &\q, \qqq(\resub{\m,\,} \q)):=\partial_\theta (-\beta f_{1RSB}(\resub{\mm,\,} \bar{q}_1,\q_0 \vert \theta) )\vert_{\theta=1} 
    \notag\\
    =&  - \dfrac{\alpha}{2} \log \left[\dfrac{\Delta_1(\qqq(\resub{\m,\,} \q))}{\Delta_1(\q)}\right] + \dfrac{\alpha \beta (\qqq(\resub{\m,\,} \q) - \q)}{2 \Delta_1(\q)} - \dfrac{\alpha \beta^2 (1-\qqq(\resub{\m,\,} \q))(\qqq(\resub{\m,\,} \q)-\q)}{2\Delta_1(\qqq(\resub{\m,\,} \q))\Delta_1(\q)} \notag \\
    &- \mathbb{E} \log \cosh \left(\beta \m+ \beta z \dfrac{\sqrt{\alpha \q}}{2 \Delta_1(\q)}\right) + \mathbb{E}_1 \left[ \dfrac{\mathbb{E}_2 \cosh g_1( \resub{\m,\,} \qqq(\resub{\m,\,} \q), \q) \log \cosh g_1(\resub{\m,\,} \qqq(\resub{\m,\,} \q), \q)}{\mathbb{E}_2 \cosh g_1(\resub{\m,\,} \qqq(\resub{\m,\,} \q), \q)}\right].
    \label{eq:K_t1}
\end{align}
Next, we study the sign of \eqref{eq:K_t1}, where $\q$ and $\qqq(\resub{\m,\,} \q)$ are the solutions of the self-consistency equations \eqref{eq:AGSq} and \eqref{eq:self2_t1}, respectively. To this purpose, it is useful to study the behaviour of the function $K(\resub{\m,\,} \q, x)$ for $x \in [0, \q]$. For $x=\q$, \resub{regardless of the value assigned to $\m$,} we have $K(\resub{\m,\,} \q, \q)=0$,
% Moreover, if we consider the limit $\qq \to \q$, we have that $K(\q, \qq)=0$. 
%Furthermore, proceeding to find the potential critical points, namely 
while the extremum of $K(\resub{\m,\,} \q,x)$ is found from 
\begin{align}
    \partial_{x} K(\resub{\m,\,} \q, x) &= \dfrac{\beta^2 \alpha x}{2\Delta_1(x)^2 } \left[ x- \mathbb{E}_1 \dfrac{\mathbb{E}_2 \cosh g_1(\resub{\m,\,} x, \q)\tanh^2 g_1(\resub{\m,\,} x, \q)}{\mathbb{E}_2 \cosh g_1(\resub{\m,\,} x, \q)} \right]=0
 \notag   \end{align}
    at 
    \begin{align}
x= \mathbb{E}_1 \dfrac{\mathbb{E}_2 \cosh g_1(\resub{\m,\,} x, \q) \tanh^2 g_1(\resub{\m,\,} x, \q)}{\mathbb{E}_2 \cosh g_1(\resub{\m,\,} x, \q)}\equiv \qqq(\resub{\m,\,} \q), 
\end{align}
where the last equality follows from Eq. \eqref{eq:self2_t1}. Given that $K(\resub{\m,\,}\q,x)$ vanishes for $x=\q$, if the extremum $x=\qqq(\resub{\m,\,} \q)$ is global in the domain considered, we must have that $K(\bar q, \tilde q_1(\q))>0$ if $x=\tilde q_1(\q)$ is a maximum and $K(\resub{\m,\,} \bar q, \tilde q_1(\q))<0$ if $x=\tilde q_1(\resub{\m},\q)$ is a minimum. Therefore, if 
\begin{align}
    \partial_{x}^2 K(\q, x) \vert_{x=\qqq(\resub{\m,\,} \q)} & = 
    -\dfrac{\beta^2 \alpha }{2\Delta_1(\qqq(\resub{\m,\,} \q))^2} \left\{ 1-\dfrac{\beta^2 \alpha}{\Delta_1(\qqq(\resub{\m,\,} \q))^2} \mathbb{E}_1 \left[ \dfrac{\mathbb{E}_2 \sech^3 g_1(\resub{\m,\,} \qqq(\resub{\m,\,} \q), \q)}{\mathbb{E}_2 \cosh g_1(\resub{\m,\,} \qqq(\resub{\m,\,} \q), \q)}\right]^2\right\}
    \label{eq:second-der_t1}
\end{align}
is positive, $K(\resub{\m,\,} \q, \qqq(\resub{\m,\,} \q))$ is negative and 
\begin{align}
  {f}_{1RSB}(\resub{\m,\,} \qqq(\resub{\m,\,} \q), \q\vert \theta)= f_{RS}(\resub{\m,\,}\q)- (\theta-1) \frac{K(\resub{\m,\,} \qqq(\resub{\m,\,} \q), \q)}{\beta}< f_{RS}(\resub{\m,\,}\q),
\end{align}
hence the RS theory becomes unstable
when the expression in the curly brackets in \eqref{eq:second-der_t1}
becomes negative i.e. for
\begin{equation}
     (1-\beta(1-\qqq(\resub{\m,\,} \q)))^2 < {\beta^2 \alpha} \mathbb{E}_1 \left[\dfrac{\mathbb{E}_2 \sech^3 g_1(\resub{\m,\,}\qqq(\resub{\m,\,} \q), \q)}{\mathbb{E}_2 \cosh g_1(\resub{\m,\,} \qqq(\resub{\m,\,} \q), \q)}\right]^2
     \label{eq:AT-hop_t1}
\end{equation}
This expression recovers the result found by Coolen in \cite{coolen2001statistical} using the de Almeida-Thouless approach \cite{de1978stability}, in the 
limit $\qqq(\resub{\m,\,} \q)\to \q$, where \eqref{eq:AT-hop_t1} reduces to
\begin{equation}
     (1-\beta(1-\q))^2 < {\alpha\beta^2} \,\mathbb{E} \cosh^{-4} \left[\beta \m \!+\! \beta z \dfrac{\sqrt{\alpha \q}}{1-\beta(1-\q)}
     \right].
     \label{eq:AT-hop_old_t1}
\end{equation}

\begin{figure}[h]
    \centering
    \includegraphics[width=15.5cm]{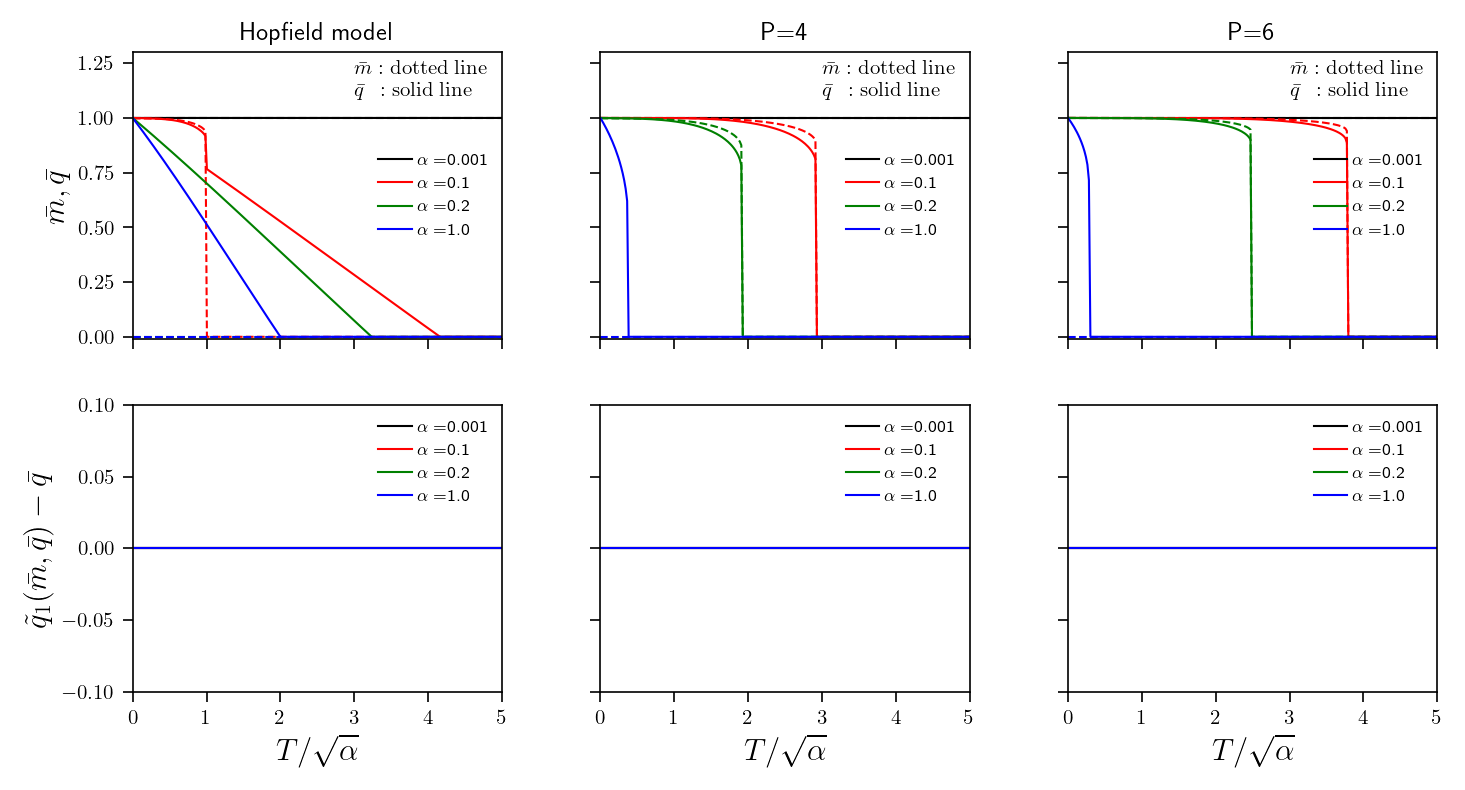}
    \caption{
    RS overlap $\q$ (top row) and the difference between $\qqq(\resub{\m,\,} \q)$ and $\q$ (bottom row) versus the scaled parameter $T/\sqrt{\alpha}$, for different values of $\alpha$ (as shown in the legend), for the Hopfield model (left) and Hebbian networks with $P$-node interactions, for $P=4$ (mid) and $P=6$ (right). %Dotted lines show the critical (scaled) temperature $T_c/\sqrt{\alpha}$ at which the RS instability occurs (lines move to the left when $\alpha$ is increased in the Hopfield model, whereas they do not depend on $\alpha$ in Hebbian networks with $P>2$).
    }
    \label{fig:diffq0qHOP}
\end{figure}

While this limit is {\it a priori}
unjustified, as
$\q$ and $\qqq(\resub{\m,\,} \q)$ should be solved from the self-consistency equations \eqref{eq:AGSq} and \eqref{eq:self2_t1}, respectively, one can check numerically 
that the solutions of these equations are virtually indistinguishable for any temperature (see Fig.   \ref{fig:diffq0qHOP}, left panel), and the resulting RS instability line is almost identical to the AT line derived in \cite{coolen2001statistical}, see Fig. \ref{fig:HOP-AT} \resub{(left panel). Small deviations can be appreciated in the retrieval region (see right panel), but these are likely due to numerical precision.}
%\resub{Indeed, the model is known to undergo a second order continuous phase transition in the order parameter $q$}. 

\begin{figure}[h]
    \includegraphics[width=15cm]{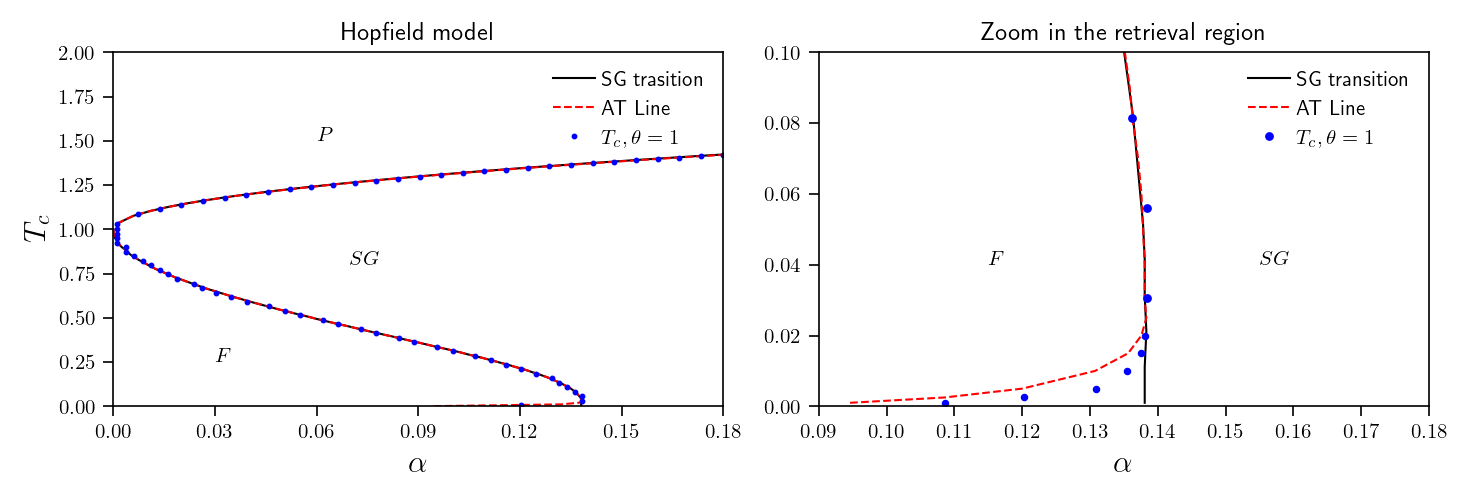}
    \caption{
RS instability line (i.e. $T_c$ versus $\alpha$) obtained via our method (blue crosses) and the AT line obtained in the limit $\qqq(\resub{\m,\,} \q)\to \q$ (red dashed curve), for the Hopfield model model.  The several branches separate the paramagnetic (P), spin-glass (SG) and retrieval (M) region.
The black curve show the critical temperature $T^\star$ at which $\q$ becomes non-zero within the RS theory, i.e. the spin-glass (SG) transition. }
    \label{fig:HOP-AT}
\end{figure}

% \iffalse 
% We anticipate, however, that such differences become more pronounced in Hebbian networks with $P$-node interactions, that 
% we will analyse in the next section, 
% which undergo a discontinuous phase transition, such that $\q_1$ differs from $\q_0$ even at the onset of the RS instability, implying $\qqq(\resub{\m,\,} \q)\neq \q$. 
% \fi
As our method 
does not rely on the assumption
$\qqq(\resub{\m,\,} \q)\to \q$, it provides a more general approach than the one originally devised by de Almeida and Thouless, that can be 
carried over to systems with discontinuous phase transitions, where $\q_1$ differs from $\q_0$ even at the onset of the RS instability, implying $\qqq(\resub{\m,\,} \q)\neq \q$.

Before concluding this section, we note that, although we have disregarded the limit 
$\theta\to 0$ as lacking physical interpretation, such limit is still well defined mathematically and one may ask what would be 
the outcome of a similar analysis carried out in this limit. We will perform such analysis in Appendix \ref{app:expt0}. 
% \iffalse
% Reassuringly, we find that such analysis generally leads to a lower temperature for the RS instability, 
% confirming that the limit $\theta\to 1$ gives the physical transition and it is therefore the relevant one. 
% Intriguingly, however, in the Hopfield model the analysis for $\theta\to 0$ gives the same instability line as the analysis for $\theta\to 1$.  
% \fi
\resub{Intriguingly, we find that the analysis for $\theta\to 0$ gives the same instability line as the analysis for $\theta\to 1$, in all the models we considered, except in the spherical $P$-spin model. Reassuringly, in the latter case, the analysis at $\theta\to 0$ leads to a lower temperature for the RS instability, 
confirming that the limit $\theta\to 1$ gives the physical transition and it is therefore the relevant one.}

\noindent

\section{Hebbian networks with $P$-node interactions}
\label{sec:dense}
In this section we consider generalizations of the Hopfield model, where neurons interact in P-tuples of even $P \geq 4$  (rather than pairwise, i.e. $P=2$). Such networks were shown to store many more patterns than the number of their nodes, so they work as {\it dense} associative memories \cite{HopKro1}. They are also known to be dual to deep neural networks \cite{Fachechi1, Albanese2021} and to exhibit information processing capabilities that are forbidden in shallow networks, such as the existence of a region in the parameter space
where they can retrieve patterns although these are overshadowed by the noise  \cite{Barra-PRLdetective}.
As before, we consider a network of $N$ interacting Ising neurons $\sigma_i\in\{1,-1\}$, with $K$ stored patterns 
$\bm \xi^\mu$ 
%\resub{whose $N$ Rademacher entries are now grouped in sets each made of $P/2$ elements.} 
%which are now vectors of $N^{P/2}$ Rademacher entries. 
and $P$-node interactions $J_{i_1\ldots i_P}=\frac{1}{P!}\sum_{\mu=1}^K \xi^\mu_{i_1}\ldots \xi^\mu_{i_P}$.
The Hamiltonian of this model can be written as 
\begin{align}
    {H_N(\bm \sigma \vert \bm \xi)= -\dfrac{N^{1-P}}{P! } \sum_{\mu=1}^K \sum_{i_1, \hdots, i_{P}} \xi^\mu_{i_1}  \hdots \xi^\mu_{i_{P}} \sigma_{i_1} \hdots \sigma_{i_{P}}}
    \label{eq:H_Dense}
\end{align}
The order parameters are still the Mattis magnetization $m$ and the two-replicas overlap $q$, as introduced in \eqref{eq:def_m}, with their RS distributions given in (\ref{eq:P_RS_Hop1}) and (\ref{eq:P_RS_Hop2}) and their 1RSB generalizations given in \eqref{eq:1RSBAss} and \eqref{eq:1RSBAssm}.
% \begin{eqnarray}
% \label{eq:order_Dense}
%          m&:=\dfrac{1}{N}\SOMMA{i=1}{N}\xi_i^{1}\sigma_i \ \ 
%          q_{12}&:=\dfrac{1}{N}\SOMMA{i=1}{N}\sigma_i^{(1)}\sigma_i^{(2)}.
% \end{eqnarray}
%
The quenched free-energy in RS assumption is (see \cite{Gardner})
\begin{align}
    -\b f_{RS}(\m, \q \vert \b, \alpha) =& \ln 2 - \dfrac{{\b}}{2}(P-1) \m^P + \dfrac{{\b}^2 \alpha}{4}(1-\q^P)- \dfrac{{\b}^2 \alpha P }{4} \q^{P-1} (1-\q)\notag \\
    &+ \mathbb{E} \ln \cosh \left( \b \dfrac{P}{2} \m^{P-1} + \b z \sqrt{\alpha \dfrac{P}{2} \q^{P-1}}\right) 
\end{align}
with $\b:=2\beta/P!$, where $\beta$ is the inverse temperature and $\alpha= \lim_{N \to \infty} K/N^{P-1}$ is the network load. $\mathbb{E}$ is the average w.r.t. the standard Gaussian random variable $z$, and $\q$ and $\m$ satisfy the self-consistency equations: 
\begin{align}
    \m= \mathbb{E} \tanh \left( \b \dfrac{P}{2} \m^{P-1} + \b \sqrt{\alpha\dfrac{P}{2}\q^{P-1}} z\right), \notag \\
    \q= \mathbb{E} \tanh^2 \left( \b \dfrac{P}{2} \m^{P-1} + \b \sqrt{\alpha\dfrac{P}{2}\q^{P-1}} z\right).
    \label{eq:self_RS_Dense}
\end{align}
On the other hand, the quenched free-energy within the 1RSB approximation (see \cite{Albanese2021}), reads as
\begin{align}
    -\b f_{1RSB}(\mm, \q_1, \q_0 \vert \b, \alpha, \theta) =& \ln 2 - \dfrac{{\b}}{2}(P-1) \mm^P + \dfrac{{\b}^2 \alpha }{4} \left[ 1-\theta \q_0^P + (\theta-1)\q_1^P\right] 
    \notag \notag \\
&-\dfrac{{\b}^2 \alpha P }{4} \q_1^{P-1} - \dfrac{{\b}^2}{4} P \alpha\left[ (\theta-1) \q^P_1 - \theta \q_0^P \right] \notag \\
    &+ \dfrac{1}{\theta} \mathbb{E}_1 \ln \mathbb{E}_2 \cosh^\theta g(\b, \alpha, \mm, \q_0, \q_1) 
    \label{eq:A1RSBDense}
\end{align}
where $\mathbb{E}_1$, $\mathbb{E}_2$ are the average w.r.t. the standard normal random variables $z^{(1)}$ and $z^{(2)}$, 
%({\em vide infra}), 
respectively, and 
\begin{align}
\label{eq:gDense}
    g(\mm, \q_1, \q_0 \vert \b, \alpha)=&\dfrac{\b P}{2} \mm^{P-1} + \b z^{(1)} \sqrt{\dfrac{P}{2} \alpha \q_0^{P-1}}+ \b z^{(2)} \sqrt{\dfrac{P}{2} \alpha \left(\q_1^{P-1}-\q_0^{P-1}\right)}.
\end{align}
\normalsize
In this approximation the self-consistency equations for the order parameter $\q_1$, $\q_0$ and $\m$ are as in \eqref{eq:self_RSBHOP}, with the argument of the hyperbolic cosine and tangent replaced by \eqref{eq:gDense}. As before, for $\theta=1$, $\q_0=\q$
and the 1RSB expression for the quenched free-energy reduces to the RS one.

From now on, we imply the dependence of the functions  \resub{on $\b$ and $\alpha$}. Our objective is to prove that for $\theta$ close but away from one, the 1RSB quenched free-energy is smaller than its replica symmetric counterpart i.e. $f_{1RSB}(\resub{\mm,\,} \q_1,\q_0|\theta) < f_{RS}(\resub{\m,\,}\q)$ above a critical value of the effective parameter $\sqrt{\alpha} \beta'$.
To this purpose, we proceed as in the Hopfield model: we expand, to the leading order in $\theta-1$, the 1RSB quenched free-energy around its RS expression, as shown in \eqref{eq:expansionHOP_t1}.
Since the self-consistency equations also depend on $\theta$, we need to expand them too. 
Following the same steps as in the Hopfield model, we can write \resub{$\mm$ as in \eqref{eq:m_t1} with $C(\m, \q,\resub{\qqq(\m, \q)})$ as in \eqref{eq:C-Dense_t1},} 
where $\qqq(\resub{\m,\,} \q)$ is the solution of the self-consistency equation 
\begin{align}
     \q_1&=\mathbb{E}_1 \left[ \dfrac{\mathbb{E}_2 \cosh {g(\resub{\m,\,}\q_1, \q)} \tanh^2 {g(\resub{\m,\,}\q_1, \q)}}{\mathbb{E}_2 \cosh {g(\resub{\m,\,}\q_1, \q)}}\right], 
     %=  \mathbb{E}_1 \left[ \dfrac{\mathbb{E}_2 \sinh {g(\q_1, \q)} \tanh {g(\q_1, \q)}}{\mathbb{E}_2 \cosh {g(\q_1, \q)}}\right]
     \label{eq:q1Dense_t1}
\end{align}
$\q_0$ as in \eqref{eq:q0-q_t1}, 
%\eqref{eq:q1-q-pre_t1}, 
with $A(\resub{\mm,\,} \q,\qqq(\m, \q))$ given in \eqref{eq:A-Dense_t1}, and $\q_1$ as given in \eqref{eq:q1-q_t1} with 
$B(\resub{\m,\,} \q,\qqq(\resub{\m,\,} \q))$ given in \eqref{eq:B-Dense_t1}.
% \iffalse
% \begin{align}
%     \q_1 &\sim \mathbb{E} \tanh^2 g(\q_1, \q_0) \notag \\
%     &+ \theta \left[ \mathbb{E}_1\mathbb{E}_2 \ln \cosh g(\q_1, \q_0) \tanh^2 g(\q_1, \q_0) \right.\notag \\
%     &\left.- \mathbb{E}_1 \left( \mathbb{E}_2 \ln \cosh g(\q_1, \q_0)  \mathbb{E}_2 \tanh^2 g(\q_1, \q_0)  \right)\right] \notag \\
%     &=:\q + \theta A(\q_1, \q_0), 
%     \end{align}
%     \normalsize
% and, for $\theta=0$, the expression is reduced as 
% \begin{align}
%     \q_1 = \mathbb{E} \tanh^2 g(\q_1, \q_0) 
% \end{align}
% which is equivalent to say that $\q_1$ satisfies the RS self-consistency equation.
%     \begin{align}
%     &\q_0 \sim \mathbb{E}_1\left( \mathbb{E}_2 \tanh g(\q, \q_0) \right)^2 \notag \\
%     &+ \theta \left\{ 2 \mathbb{E}_1 \left[ \mathbb{E}_2 \tanh g(\q, \q_0) \mathbb{E}_2 \ln \cosh g(\q, \q_0) \tanh g(\q, \q_0) \right] \right.\notag \\
%     &\left.- 2 \mathbb{E}_1 \left[ \left( \mathbb{E}_2 \tanh g(\q, \q_0) \right)^2 \mathbb{E}_2 \ln \cosh g(\q, \q_0) \right]\right\} \notag \\
%     &=: \mathbb{E}_1\left( \mathbb{E}_2 \tanh g(\q, \q_0) \right)^2 + \theta B(\q, \q_0).
% \end{align}
% Also in this case, for $\theta=0$ we reach the following self-consistency equation
% \begin{align}
%      \q_0&=\mathbb{E}_1\left( \mathbb{E}_2 \tanh g(\q, \q_0) \right)^2 
%      \label{eq:q0Dense}
% \end{align}
% which is dependent only on $\q$.
% We neglect $\m$ since its distribution is independent from $\theta$.
% \fi
%%%%%%%%%%%%%%%%
%%%%%%%
%
With the above expressions in hand, we can now calculate the derivative of $f_{\rm 1RSB}$ w.r.t. $\theta$ when $\theta=1$, as needed in \eqref{eq:expansionHOP_t1}
\begin{align}
\label{eq:KDense_t1}
    &K(\resub{\m,\,} \qqq(\resub{\m,\,} \q), \q):=\partial_\theta (-\b f_{\rm 1RSB}(\resub{\mm,\,}\q_1, \q_0|\theta) )\vert_{\theta=1} 
    \notag\\
    &=-\dfrac{{\b}^2 \alpha}{4} (P-1)[(\tilde{q}_1(\resub{\m,\,} \q))^P- \q^P] -\dfrac{{\b}^2 \alpha }{4} P[(\tilde{q}_1(\resub{\m,\,} \q))^{P-1} - \q^{P-1}] \notag \\
    &- \mathbb{E} \ln \cosh \left( \b \sqrt{\dfrac{P}{2} \q^{P-1}}z + \b \m^{P-1} \right) + \mathbb{E}_1 \left[ \dfrac{\mathbb{E}_2 \cosh g(\resub{\m,\,}\qqq(\resub{\m,\,} \q), \q) \log \cosh g(\resub{\m,\,}\qqq(\resub{\m,\,} \q), \q) }{\mathbb{E}_2 \cosh g(\resub{\m,\,}\qqq(\resub{\m,\,} \q), \q) }\right]
\end{align}
Again, we have that $K(\resub{\m,\,} \q, \q)=0$, \resub{regardless of the value assigned to $\m$,} (this follows from the fact that for $\theta=1$, $\q$ is an extremum of the free-energy). Next, we inspect the sign of $K(\resub{\m,\,}\qqq(\resub{\m,\,} \q), \q)$. To this purpose, we study $K(\resub{\m,\,}\q, x)$ for $x \in [0, \q]$ and locate its extrema, which are found from 
\begin{align}
    \partial_{x} K(\resub{\m,\,}\q, x)=& -\dfrac{{\b}^2 \alpha}{4} (P-1) P x^{P-2}\left[x-   \mathbb{E}_1 \left[ \dfrac{\mathbb{E}_2 \sinh g(\resub{\m,\,}\q, x) \tanh g(\resub{\m,\,}\q, x) }{\mathbb{E}_2 \cosh g(\resub{\m,\,}\q, x) }\right]\right]=0
    \end{align}
    as 
\begin{align} x = \mathbb{E}_1 \left[ \dfrac{\mathbb{E}_2 \cosh {g(x, \q)} \tanh^2 {g(x, \q)}}{\mathbb{E}_2 \cosh {g(x, \q)}}\right] \equiv \qqq(\resub{\m,\,} \q)
\end{align}
where the last equality follows from \eqref{eq:q1Dense_t1}. 
Under the assumption that 
the extremum $x=\qqq(\resub{\m,\,} \q)$ is global in the domain considered and reasoning as in the 
Hopfield case, we have that 
$K(\resub{\m,\,}\qqq(\resub{\m,\,} \q), \q)>0$ if $x=\tilde q_1(\resub{\m,\,} \q)$ is a maximum and $K(\resub{\m,\,} \qqq(\resub{\m,\,} \q), \q)<0$
if it is a minimum. In particular, if 
\begin{align}
    \partial^2_{x} K(\resub{\m,\,}\q,x)\vert_{x=\qqq(\resub{\m,\,} \q)}=& -\dfrac{{\b}^2 \alpha (P-1) P}{4}  (\tilde{q}_1(\resub{\m,\,}\q))^{P-2} \notag \\
    &\cdot\left\{ 1-\dfrac{{\b}^2 \alpha}{2} (P-1) P (\tilde{q}_1(\resub{\m,\,} \q))^{P-2} \mathbb{E}_1 \left[ \dfrac{\mathbb{E}_2 \textnormal{sech}^3 g(\resub{\m,\,}\qqq(\resub{\m,\,} \q), \q)  }{\mathbb{E}_2 \cosh g(\resub{\m,\,}\qqq(\resub{\m,\,} \q), \q)}\right]\right\}
    \label{eq:K2-Dense}
\end{align}
is positive, $K(\resub{\m,\,}\qqq(\resub{\m,\,} \q), \q)<0$ and $f_{1RSB}<f_{RS}$. This happens when the expression in the curly brackets of the equation above is negative, i.e. when the parameter $\alpha {\b}^2$ satisfies the inequality
\begin{equation}\label{eq:result_PHOP}
\dfrac{{\b}^2 \alpha}{2} (P-1) P (\tilde{q}_1(\resub{\m,\,}\q))^{P-2} \mathbb{E}_1 \left[ \dfrac{\mathbb{E}_2 \textnormal{sech}^3 g(\resub{\m,\,}\q, \qqq(\resub{\m,\,} \q) ) }{\mathbb{E}_2 \cosh g(\resub{\m,\,} \q, \qqq(\resub{\m,\,} \q))}\right] >1.
\end{equation}
As noted in \cite{Albanese2021,glassy}, Hebbian networks with $P$-node interactions are equivalent to Ising $P$-spin models  
under a suitable rescaling of the temperature $\b \sqrt{\alpha}\to \beta'$. With this rescaling, \eqref{eq:result_PHOP} retrieves indeed the RS instability line of the Ising P-spin model, that we have for completeness derived in Appendix \ref{app:spin-glasses}, using 
our method (see eq. \eqref{eq:ATPspin}). In the limit $\qqq(\resub{\m,\,} \q)\to \q$,
\eqref{eq:ATPspin} retrieves the AT line of the Ising $P$-spin model \cite{gardner1985spin}.
% \iffalse, however  we note that 
% for the Ising $P$-spin model and for Hebbian networks with 
% $P$-node interactions, $\qqq(\resub{\m,\,} \q)$ differs from $\q$, %with deviations getting more pronounced as $P$ is increased, 
% hence this limit cannot be justified. 
% This is due to the discontinuous nature (in the order parameter $q$)
% of the transition from RS to 1RSB, in these models.
% \fi
In Fig. \ref{fig:diffq0qHOP} (mid and right panels) we plot the difference between $\qqq(\resub{\m,\,} \q)$ and $\q$ for Hebbian networks with $P$-node interactions
(obtained solving numerically the self-consistency equations %\eqref{eq:self2_t1} and \eqref{eq:AGSq} for the Hopfield model, and equations  
\eqref{eq:q1Dense_t1} and \eqref{eq:self_RS_Dense}) as a function of the scaled parameter $T/\sqrt{\alpha}$, for different values of $\alpha$. 
% \iffalse
% At difference with the Hopfield model (left) in networks with multi-node interactions, differences can be appreciated at the onset of the RS instability (results shown for $P=4$ and $P=6$).
% \fi
\resub{As for the Hopfield model, we find that $\qqq(\resub{\m,\,} \q)$ is indistinguishable from $\q$, hence the limit $\qqq(\resub{\m,\,} \q)\to \q$ can be justified {\it a posteriori}. }

\begin{figure}[h]
    \centering
    \includegraphics[width=15.5cm, height=4.5cm]{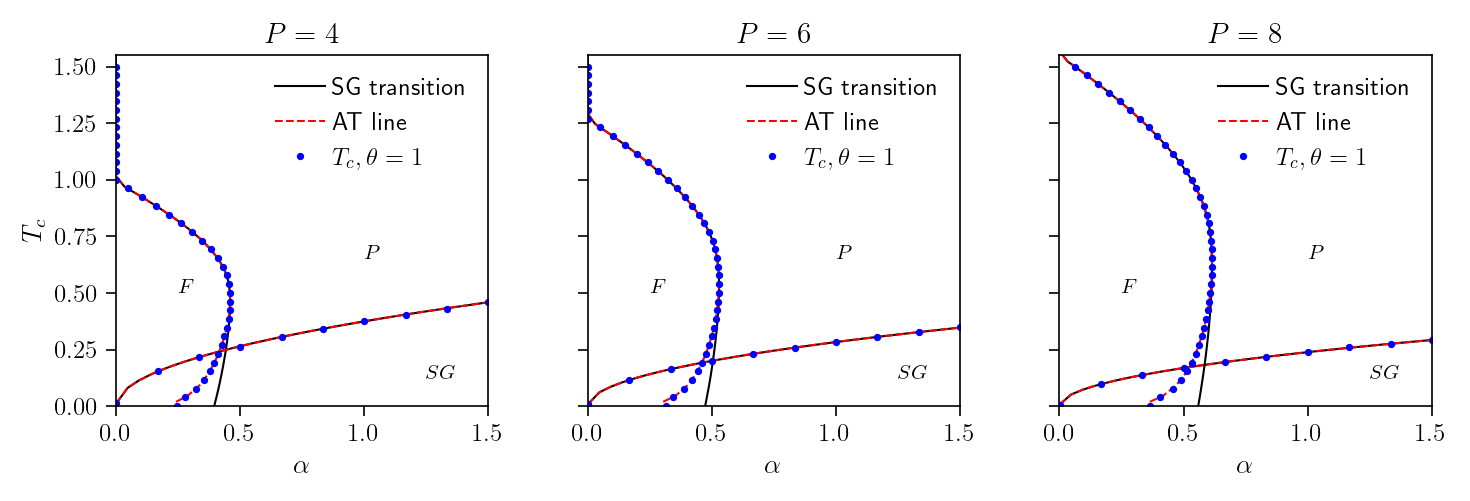}
    \caption{
RS instability line (i.e. $T_c$ versus $\alpha$) obtained via our method, i.e. by expanding the free-energy around $\theta=1$ (blue dots) and the AT line obtained in the limit $\qqq(\resub{\m,\,} \q)\to \q$ (red curve), for the Hebbian network with $P$-node interactions, with $P=2, 4, 6$ from left to right. 
The black curves show the critical temperature $T^\star$ at which $\q$ becomes non-zero within the RS theory, i.e. the spin-glass (SG) transition. 
%\resub{For comparison, we also show the RS instability line obtained by expanding around $\theta=0$ (green circles), as shown in Appendix \ref{app:}}.
}
    \label{fig:Dense-AT}
\end{figure}

In Fig. \ref{fig:Dense-AT} we show the RS instability lines 
resulting from our method and the classic AT line obtained in the limit $\qqq(\resub{\m,\,} \q)\rightarrow \q$, for different values of $P$. 
% \iffalse
% Deviations can be appreciated at $P=4$ and increase for higher values of $P$.
% \fi
% \resub{
The two lines coincides for all values of $P$. As explained earlier, we could have expanded the free-energy around $\theta=0$ (as opposed to $\theta=1$). In Appendix \ref{app:expt0} we show that such analysis leads to the same line.
%For comparison, we report in the same figure the critical line resulting from such analysis, which, intriguingly, coincides with the line obtained at $\theta=1$. }

%(see derivation in the Appendix, exactly as predicted by the decomposition theorem of Dense Hebbian Network \cite{Albanese2021}. 
%\linda{Moreover, in the limit $\qq \to \q$ we reach, a part for a constant, the expression found in \cite{bardina2004} via AT's approach for $P$-spin glass model.}

\section{Discussion}
In this work we proposed a simple and systematic method to derive the critical line in the parameter space $(\alpha, \beta)$, below which the 1RSB expression for the free-energy is smaller than the RS expression, in Hebbian neural networks. The same analysis for spin-glass models is carried out in appendix \ref{app:spin-glasses}. For the Hopfield model, our approach recovers the critical line obtained by Coolen using the AT approach \cite{coolen2001statistical} as a special limit. Similarly, we recover the known AT lines of all the spin-glass models considered in the appendix, in the same limit, showing that our method provides a generalization of the approach originally devised by de Almeida and Thouless. Owing to its simplicity, our method allows for straightforward application to Hebbian networks with multi-node interactions, for which the AT-line was unknown.

The key idea of our method is to regard the 1RSB theory, which assumes two delta-peaks in 
the overlap distribution $P(q)$, located at $\q_1$ and $\q_0<\q_1$, with weights $1-\theta$ and $\theta$, respectively,
as departing continuously from the RS theory, which assumes only one peak at $\q$. 
This leads us to assume that at the onset of the RS instability, the 1RSB overlap distribution is dominated by one peak, so that either $\theta\to 0$ or $\theta\to 1$. Then, consistency with the RS theory requires either $\lim_{\theta\to 0}\q_1\!=\!\q$ 
or $\lim_{\theta\to 1}\q_0\!=\!\q$. 
\iffalse
As $\q_1$ and $\q_0$ are determined from the self-consistency equations of the 1RSB theory, for each value of $\theta$, it is straightfoward to determine 
which scenario applies. 
Noting that for the Hopfield model and Hebbian networks with $P$-node interactions, \resub{one has $\q_0\!=\!\q$ for $\theta\!=\!1$}, 
\fi
As the physical interpretation of RS breaking 
suggests that the mutual overlap between 1RSB states should be equal to the self-overlap of the RS states, we regard the 1RSB theory  
as a continuous variation of the RS theory, when the parameter $\theta$ is decreased from one.  
Crucially, we do not make any assumption on the location of the peaks of the 1RSB theory, which are fixed by the 1RSB self-consistency equations. We then  compare the 1RSB and the RS free-energies when $\theta$ (the only free-parameter in our analysis) is close to one, by performing  simple expansions to linear orders in $\theta-1$. In doing so, we solely require that $f_{1RSB}, \q_1, \q_0$ \resub{and $\mm$} are differentiable up to the first order in a neighborhood of $\theta\!=\!1$ and that the derivative of $f_{1RSB}$ exists at $\theta=1$.

Although our method is similar in spirit to the one introduced by Toninelli in \cite{toninelli2002almeida}, there is a crucial difference, in that the latter relies 
on the assumption $\q_1\to \q$, which is, in our view, unjustified a priori. In fact, while $\q_0=\q$ for $\theta=1$, $\q_1$ may differ from $\q$, even in the limit $\theta\to 1$. This consideration also leads to a departure of our approach from the method originally devised by de Almeida and Thouless, which relies  
on a variation of the RS free-energy as the {\it order parameters} are varied continuously around their RS values. In contrast, we study the variation of the RS free-energy as the {\it statistical weight} of the order parameters is varied continuously (rather than the actual value of the order parameters).
%noting that $\q_0$ cannot be made arbitrarily close to $\q$ within a 1RSB theory.   
This approach allows us to determine 
the instability line of the RS theory also in spin-glass models 
which exhibit a discontinuous phase transition.
%, for which $\q_1\!\neq\! \q_0$ even at $T_c$. 
As a prototypical example of this class of models, we consider in Appendix \ref{app:spin-glasses} the spherical $P$-spin model \cite{Crisanti-RSB}. 
\iffalse
For this model, the 1RSB self-consistency equations show that $\lim_{\theta\to 1}\q_0\!=\!\q$ and the RS instability line can 
be found by comparing the 1RSB and the RS free-energies when $\theta$ is close to one. 
Although $\lim_{\theta\to 0}\q_1\!=\! \q$ {\it as well} in this model, one can show that 
a similar analysis for $\theta$ close to zero would give a lower temperature for the RS instability, hence 
the instability that occurs at higher temperature (i.e. for $\theta\!\simeq\! 1$) is the physical one.
\fi

In conclusion, in this work we have presented a new method to find the instability line of the RS approximation. A compelling advantage of our method, when compared to the approach by de Almeida and Thouless \cite{de1978stability}, is that it does not require to compute the full eigenspectrum 
of the Hessian of the quadratic fluctuations of the free-energy around its RS value and it does not rely on the availability 
of an ``ansatz-free" expression for the free-energy. This makes the computations easier and affordable also for neural networks. The method still requires the availability of 
explicit expressions for the RS and 1RSB 
free-energies,
which however can be computed using different
techniques (e.g. Guerra’s interpolation) in addition to the replica trick. 
Our approach can in principle be extended to compute the stability of the $k$-RSB solution, expanding the corresponding 
$k\!+\!1$-RSB free-energy, provided one has an explicit expression for the two. 
\resub{For example, it would be interesting to see whether the well-known transition from $1$RSB to full-RSB occurring at the Gardner temperature \cite{gardner1985spin} in the Ising $P$-spin model can be recovered within our approach, by studying the stability of the $2$RSB solution. In addition, in recent years there has been a boost of renewed interest in mixed $P$-spherical models, as introduced in \cite{AnnibaleGualdiCavagna04, CrisantiLeuzzi04}, as they have been shown to display unexpected 
dynamical behaviour \cite{FolenaFranzTersenghi20} and new types of spin-glass phases \cite{CrisantiLeuzzi05}, \cite{TonoloNieddaGradenigo} as well as for their relevance to the modelling of random lasers \cite{Antenucci15, Antenucci15b, Antenucci15c, Antenucci16, Antenucci21}. These models similarly display transitions from $1$RSB to full-RSB and would be good lab systems to test extensions of our theory.}
%Such extension could be a future outlook of this work. 
Another interesting avenue for future work would be a generalization of the $1$-RSB scheme, which does not assume the   
Mattis magnetization to be self-averaging. Preliminary work in \cite{AABO-JPA2020} suggests that the RS assumption is not the right approximation for Mattis magnetization. 
%\old{Finding the correct assumption would allow us to generalize the method presented in this manuscript to analyse the instability of the RS solution in the retrieval phase. IN or OUT?}

\resub{Finally, an attractive perspective would be to apply our approach to predict the onset of ergodicity breaking in systems with {\it sparse} interactions. In such systems, the free-energy is typically expressed, already at the simplest RS level, in terms of order-parameter {\it functions} to be determined self-consistently.  
At 1RSB level, recursive equations for functional distributions of such functions must be solved. Working out the fluctuations of ansatz-free free-energies around their RS value, as it would be required by the de Almeida and Thouless approach would be unfeasible and no similar approach has been proposed to date. An alternative approach has been devised in \cite{Thouless86}, for the Bethe lattice with regular degrees, however it strongly relies on the assumptions of homogeneity in the network nodes and large degrees.  
We envisage that, owing to its simplicity, our method may carry over to more general sparse systems, where it would require an expansion of the functional 1RSB self-consistency equation to linear orders around the RS equation, which should be feasible.}

\acknowledgments

%The authors gratefully acknowledge useful conversations with Andrea Alessandrelli. 
This work is supported by Ministero degli Affari Esteri e della Cooperazione Internazionale (MAECI) via the BULBUL grant (Italy-Israel), CUP Project n. F85F21006230001.
\newline
L.A. acknowledges E. Zegna Founder's Scholarship, UMI (Unione Matematica Italiana), INdAM –GNFM Project (CUP E53C22001930001) and PRIN grant  {\em Stochastic Methods for Complex Systems} n. 2017JFFHS for financial support and the Department of Mathematics at King's College London for the kind hospitality.
L.A. and A.B. acknowledge INDAM (Istituto Nazionale d'Alta Matematica) and the PRIN grant {\em Statistical Mechanics of Learning Machines} n. 20229T9EAT for support.
All the authors acknowledge the stimulating research environment provided by the Alan Turing Institute's Theory and Methods Challenge Fortnights event ``Physics-informed Machine Learning".

%\bibliographystyle{abbrv}
%\bibliography{AABAPHYSA.bib} 

\appendix

\section{Applications to Spin-glasses models}
\label{app:spin-glasses}

In this appendix, we derive the critical line for the instability of the RS theory for three archetypical spin-glass models, namely the SK model, the Ising $P$-spin and the spherical $P$-spin model, using the technique developed in the main text. In all cases, we will recover the AT lines known in the literature for the three models, 
%given in \cite{de1978stability}, \cite{} and \cite{} for the SK model, the Ising $P$-spin and the spherical $P$-spin model, respectively,
in a specific limit, 
confirming the validity and higher generality of our approach.

% \iffalse
% Beyond obtaining results that are in plain accordance with the Literature (thus confirming the soundness of our approach by an indirect proof), it is worth recalling that these two types of spin glasses can be seen as the overloaded limits of the Hopfield and the dense Hebbian neural networks respectively \cite{Albanese2021,glassy} (under stuiable noise rescaling), hence it is rather natural to recover these known expressions for the AT lines in these cases, that will be achieved separately for $P=2$ (i.e. the Sherrington-Kirkpatrick model, SK) and even $P>2$ (the  P-spin-glass) in the following.
% \fi 

\subsection{Sherrington-Kirkpatrick Model}
\label{app:SK}
The Sherrington-Kirkpatrick (SK) model \cite{sherrington1975solvable} is a system of $N$ Ising spins $\sigma_i \in \{ \pm 1 \}$ interacting via symmetric pairwise interactions 
$J_{ij}$ which are i.i.d. Gaussian variables with zero average and variance $J^2/N$. The Hamiltonian of the model is 
\begin{align}
    H_N(\bm \sigma \vert J) := - \dfrac{1}{2} \sum_{i,j}^{N,N} J_{ij} \si \sigma_j
\end{align}
and the order parameter is the two-replica overlap {$q$} as defined in \eqref{eq:def_m}.
The quenched free-energy in RS assumption is \cite{sherrington1975solvable}
\begin{align}
    -\beta f_{RS}(\q \vert \beta, J) = \ln 2 + \dfrac{\beta^2 J^2}{4} (1-\q)^2 + \mathbb{E} \left[ \ln \cosh \left( \beta J \sqrt{\q} z\right)\right]
\end{align}
where $\mathbb{E}$ is the average w.r.t. the Gaussian variable $z$ and the order parameter $\q$ fulfills the self-consistency equation  
\begin{align}
    \q= \mathbb{E} \tanh^2 \left( \beta J \sqrt{\q} z\right).
    \label{eq:self_RS}
\end{align}
The 1RSB approximation of the quenched free-energy is (see e.g. \cite{nishimori2001statistical})
\begin{align}
    -\beta f_{1RSB}(\q_1, \q_0 \vert \beta, J, \theta) =& \ln 2 + \dfrac{\beta^2 J^2}{2} (1-\q_1)+\dfrac{1}{\theta} \mathbb{E}_1 \left[ \ln \mathbb{E}_2 \cosh^\theta g(\beta, J , \q_1, \q_0 )\right]\notag \\
        &-\dfrac{\beta^2 J^2}{4} \left[1-\q_1^2 +\theta (\q_1^2-\q_0^2) \right],
    \label{eq:A_1RSB}
\end{align}
where 
\begin{align}
g(\q_1, \q_0 \vert \beta, J)=\beta J \sqrt{\q_1-\q_0} z^{(2)} + \beta J \sqrt{\q_0} z^{(1)}
\label{eq:gSK}
\end{align}
and 
$\mathbb{E}_1$, $\mathbb{E}_2$ are the average w.r.t. the standard Gaussian variables $z^{(1)}$ and $z^{(2)}$, respectively. From now on, we imply the dependence of $g$, $f_{RS}$ and $f_{1RSB}$ on $\beta$ and $J$.  The order parameters $\bar{q}_0$ and  $\bar{q}_1$ fulfill the following self-consistency equations
\begin{align}
    \q_1 &= \mathbb{E}_1 \left[ \dfrac{\mathbb{E}_2 \cosh^\theta g(\q_1, \q_0) \tanh^2 g(\q_1, \q_0)}{\mathbb{E}_2 \cosh^\theta g(\q_1, \q_0)}\right], \notag \\
    \q_0 &= \mathbb{E}_1 \left[ \dfrac{\mathbb{E}_2 \cosh^\theta g(\q_1, \q_0) \tanh g(\q_1, \q_0)}{\mathbb{E}_2 \cosh^\theta g(\q_1, \q_0)}\right]^2.
\end{align}
%

%\subsubsection{$\theta=1$ expansion}

Noting that $\lim_{\theta\to 1}\q_0=\q$ and 
$f_{1RSB} (\q_1, \q_0 \vert \theta=1)= f_{RS} (\q)$, our objective is to expand the 1RSB quenched free-energy for $\theta\simeq 1$.
To this purpose, we expand the self-consistency equations for $\q_0$ and $\q_1$ to linear order in $\theta-1$. Proceding as in the Hopfield model, we get  
\begin{align}
    \label{eq:q0exp_SK}
    \q_0 &= \q + (\theta-1) A(\q, \qqq( \q))\\
    \q_1 &= \qqq(\q) + (\theta-1) B(\q, \qqq( \q))
    \label{eq:q1exp_SK}
\end{align}
where $\qqq(\q)$ solves the self-consistency equation 
\begin{align}
    \q_1 = \mathbb{E}_1 \left\{ \dfrac{\mathbb{E}_2 \cosh g(\q_1, \q) \tanh^2 g(\q_1, \q) }{\mathbb{E}_2 \cosh g(\q_1, \q) }\right\}
    \label{eq:qqq_q_SK}
\end{align} 
and $A(\q_0, \q_1)$ and $ B(\q_0, \q_1) $ are given in \eqref{eq:A_Pspin_t1_SK} and \eqref{eq:B_Pspin_t1_SK}, respectively. Next, we derive w.r.t. $\theta$ the 1RSB free-energy \eqref{eq:A_1RSB} where we replace $\q_0$ and $\q_1$ with \eqref{eq:q0exp_SK}, \eqref{eq:q1exp_SK}, obtaining
\begin{align}
    \partial_\theta(-\beta f_{1RSB}(\q_1, \q_0, \vert \theta)) =& -\dfrac{{\beta}^2 J^2}{4} \left[(\tilde{q}_1(\q))^2 - \q^2\right] - \dfrac{1}{\theta^2} \mathbb{E}_1 \ln \mathbb{E}_2 \cosh^\theta g(\qqq(\q), \q) \notag \\
    &- \dfrac{{\beta}^2 J^2}{2}  B(\qqq( \q), \q) \q + \dfrac{1}{\theta} \mathbb{E}_1 \left[ \dfrac{\mathbb{E}_2 \cosh^\theta g(\qqq( \q), \q) \log \cosh  g(\qqq( \q), \q)}{\mathbb{E}_2 \cosh^\theta g(\qqq( \q), \q) }\right] \notag \\
    &+ \dfrac{{\beta}^2 J^2}{2}  B(\qqq(\q), \q) \theta  \mathbb{E}_1 \left[ \dfrac{\mathbb{E}_2 \cosh^\theta g(\qqq( \q), \q) \tanh g(\qqq(\q), \q) }{\mathbb{E}_2 \cosh^\theta g(\qqq( \q), \q) }\right]^2,
\end{align}
which for $\theta=1$, using \eqref{eq:self_RS} and
similar manipulations to those used in \eqref{eq:q-lim-theta1}, evaluates to
\begin{align}
    K(\tilde{q}_1(\q), \q):=  \partial_\theta(-\beta f_{1RSB}&(\q_1, \q_0, \vert \theta)) \vert_{\theta=1} =-\dfrac{{\beta}^2 J^2}{4} \left[(\tilde{q}_1(\q))^2- \q^2\right] -\dfrac{{\beta}^2 J^2}{2} (\tilde{q}_1(\q)- \q) \notag \\
    &- \mathbb{E} \ln \cosh \left( \beta \sqrt{ \q} z\right) + \mathbb{E}_1 \left[ \dfrac{\mathbb{E}_2 \cosh g(\qqq( \q), \q)  \log \cosh g(\qqq( \q), \q)  }{\mathbb{E}_2 \cosh g(\qqq( \q), \q)  }\right].
    \label{eq:K_SK_t1}
\end{align}
%As before, for $\tilde{q}_1(\q) \to \q$, $K(\q, \q)=0$. 
We then study the sign of \eqref{eq:K_SK_t1}, where $\q$ and $\qqq(\q)$ are the solutions of the self-consistency equations \eqref{eq:self_RS} and \eqref{eq:qqq_q_SK}, respectively. To this purpose, it is useful to study the behaviour of the function $K(\q, x)$ for $x \in [0, \q]$. For $x=\q$, we have $K(\q, \q)=0$, while the extremum of $K(\q, x)$ is 
found from 
\begin{align}
    \partial_x K(x, \q) 
    % =& -\dfrac{{\b}^2 J^2}{4} (P-1) P x^{P-1}-\dfrac{{\b}^2 J^2}{4} (P-1)Px^{P-2} \notag \\
    % &+\dfrac{\b J P(P-1)x^{P-2}}{4\sqrt{P/2\  x^{P-1}}} \mathbb{E}_1\left[ \dfrac{\mathbb{E}_2\left( (\sinh g \log \cosh g + \sinh g )z^{(2)}\right)}{\mathbb{E}_2 \cosh g }\right] \notag \\
    % &- \dfrac{{\b}^2 J^2}{4} (P-1) P x^{P-2} \mathbb{E}_1 \left[ \dfrac{\mathbb{E}_2 \cosh g \log \cosh g }{\mathbb{E}_2 \cosh g }\right] \notag \\
    &=-\dfrac{{\beta}^2 J^2}{2} x + \dfrac{{\beta}^2 J^2}{2}  \mathbb{E}_1 \left[ \dfrac{\mathbb{E}_2 \sinh g(x, \q)  \tanh g(x, \q)  }{\mathbb{E}_2 \cosh g(x, \q)  }\right] =0,
\end{align}
%Therefore, we state that
as 
\begin{align}
   %\partial_x K(x, \q) = 0 \Leftrightarrow
   x= \mathbb{E}_1 \left[ \dfrac{\mathbb{E}_2 \sinh g(x, \q)  \tanh g(x, \q)  }{\mathbb{E}_2 \cosh g(x, \q)  }\right] \equiv \tilde{q}_1(\q)
\end{align}
via Eq. \eqref{eq:qqq_q_SK}. Given that $K(\q,x)$ vanishes for $x=\q$, if the extremum $x=\qqq( \q)$ is global in the domain considered, we must have that $K(\qqq( \q), \q) >0$ if $x=\tilde q_1(\q)$ is a maximum and $K(\qqq(\q), \q) <0$ if $x=\tilde q_1(\q)$ is a minimum. Therefore, if 
% The second derivative w.r.t. $x$, instead, is
% \begin{align}
%     \partial_{x^2} K(x, \q) =& - \dfrac{{\b}^2 J^2}{4} (P-1)^2 P x^{P-2} + \dfrac{{\b}^2 J^2}{4} (P-1) (P-2) P \tilde{q}_1 x^{P-3} \notag \\
%     &+ 2 \left( \dfrac{{\b}^2 J^2}{4} (P-1) P \right)^2 x^{2P-4}  - 2 \left( \dfrac{{\b}^2 J^2}{4} (P-1) P \right)^2 x^{2P-4}  \mathbb{E}_1 \left[ \dfrac{\mathbb{E}_2 \sinh g \tanh^3 g }{\mathbb{E}_2 \cosh g }\right] \notag \\
%     &- 4 \left( \dfrac{{\b}^2 J^2}{4} (P-1) P \right)^2 x^{2P-4} \tilde{q}_1 , 
% \end{align}
% which means, when $x=\tilde{q}_1$, that
\begin{align}
    \partial_x^2 K(x, \q)\vert_{x=\tilde{q}_1(\q)} =& -\dfrac{{\beta}^2 J^2}{2}  \left( 1-\dfrac{{\beta}^2 J^2}{2} \mathbb{E}_1 \left[ \dfrac{\mathbb{E}_2 \textnormal{sech}^3 g(\qqq( \q), \q)   }{\mathbb{E}_2 \cosh g(\qqq( \q), \q)  }\right]\right)
\end{align}
is positive, $K( \q, \qqq( \q))$ is negative, so the RS theory becomes unstable for 
%This occurs for 
\begin{align}
    1-\dfrac{{\beta}^2 J^2}{2} \mathbb{E}_1 \left[ \dfrac{\mathbb{E}_2 \textnormal{sech}^3\  g(\qqq(\q), \q) }{\mathbb{E}_2 \cosh g(\qqq( \q), \q)  }\right] < 0.
    \label{eq:ATt1_SK}
\end{align}

We note that for $\qqq( \q) \to \q$, 
%and using equation \eqref{eq:relation}, 
such condition retrieves the well-known AT line \cite{de1978stability}
\begin{align}
      1 - {\beta^2 J^2} \mathbb{E}\left(  \dfrac{1}{\cosh^4\left( \beta J \sqrt{\q} z\right)}\right)<0.
    \label{eq:atSK}
\end{align}

By numerically solving the self-consistency equations %(\ref{eq:AGSq}) 
\eqref{eq:self_RS}
and (\ref{eq:qqq_q_SK}), we can 
verify that %, up to numerical precision, 
$\qqq( \q)=\q$ for all temperatures, hence this limit can be justified {\it a posteriori} for the SK model (see Fig. \ref{fig:diffq0qSK}, left panel). \resub{We anticipate that this remains true for the Ising $P$-spin model analysed in the next section.}
\begin{figure}[t]
    \centering
    \includegraphics[width=\textwidth, height=0.42\textwidth]{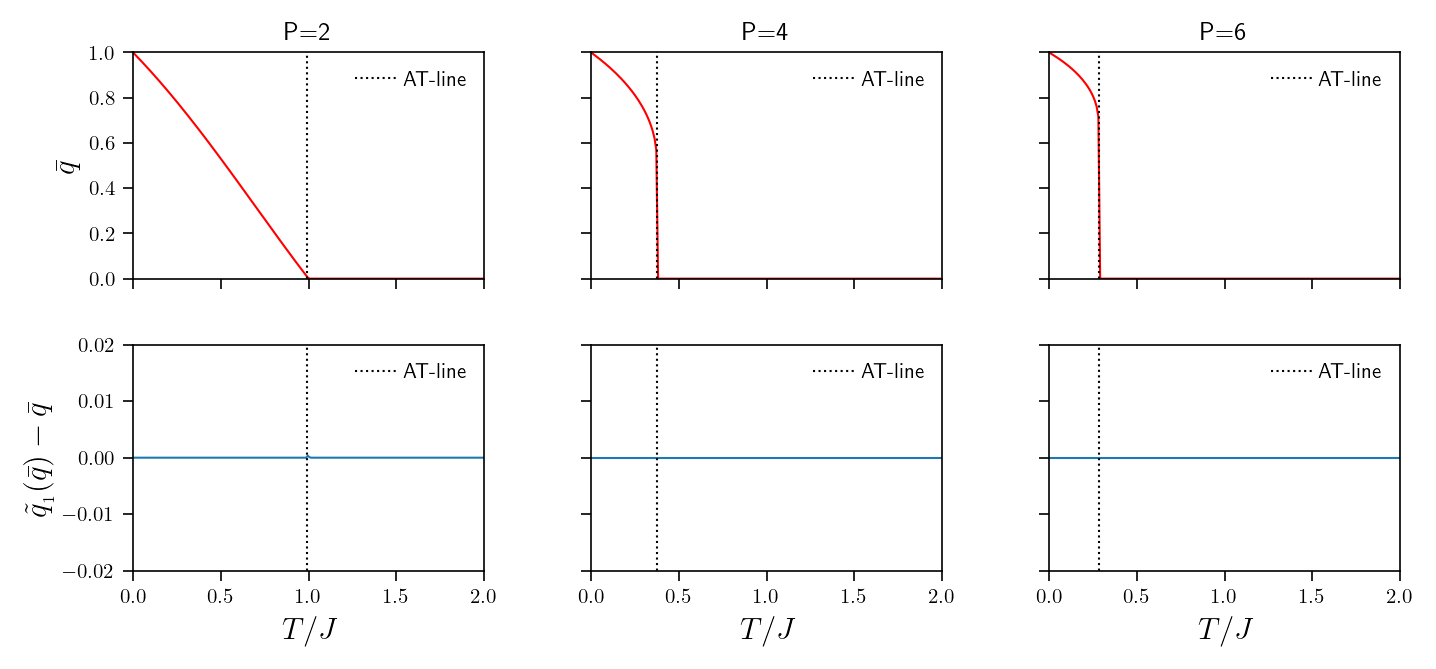}
    \caption{RS overlap $\q$ (top) and difference between $\q_1(\q)$ and $\q$ (bottom) versus the ratio $T/J$, for the SK model (left) and the Ising $P$-spin model, with $P=4$ (mid) and $P=6$ (right panel). The dotted vertical line marks the onset of the RS instability.}
    \label{fig:diffq0qSK}
\end{figure}

\subsection{The Ising P-spin model}
\label{app:PSpin}
In this section, we consider a system of $N$ Ising spins $\sigma_i=\pm 1$, $i=1, \hdots, N$ governed by the Hamiltonian 
\begin{align}
    H_N(\bm \sigma\vert J) &= -\dfrac{1}{N^{P-1}P!} \sum_{i_1, \hdots , i_P=1, \hdots , 1}^{N, \hdots , N} J_{i_1, \hdots, i_P} \sigma_{i_1} \hdots \sigma_{i_P}
    \label{eq:HamiltonianPspin}
\end{align}
where $J_{i_1, \hdots, i_P}$ are Gaussian i.i.d. variables, $J_{i_1, \hdots, i_P} \sim \mathcal{N}(0,J^2)$.
As in the previous cases, the order parameter of the model is the two-replica overlap $q$ defined in \eqref{eq:def_m}.
The quenched free-energy in RS assumption, at inverse temperature $\beta$ reads as \cite{gardner1985spin}
\begin{align}
    -\b f_{RS}(\q \vert \beta', J) =& \ln 2 + \dfrac{{\b}^2 J^2}{4}[1-P \q^{P-1} + (P-1)\q^P] + \mathbb{E} \ln \cosh \left( \b J z \sqrt{ \dfrac{P}{2} \q^{P-1}}\right)
\end{align}
where $\b= 2\beta/P!$, 
$\mathbb{E}$ is the average w.r.t. the Gaussian variable $z$ and $\q$ fulfills the following self-consistency equation: 
\begin{align}
    \q= \mathbb{E} \tanh^2  \left( \b J z \sqrt{ \dfrac{P}{2} \q^{P-1}}\right).
    \label{eq:self_RS_PSpinHOP}
\end{align}
The 1RSB expression of the quenched free-energy in the thermodynamic limit is \cite{gardner1985spin}
\begin{align}
    -\b f_{1RSB}(\q_1, \q_0 \vert \beta', J, \theta) =& \ln 2 + \dfrac{{\b}^2 J^2}{4} \left[ 1-P\q_1^{P-1} + (P-1) \q_1^P \right] \notag \\
    &+ \dfrac{1}{\theta} \mathbb{E}_1 \ln \mathbb{E}_2 \cosh^\theta g (\beta', J, \q_1, \q_0)  - \dfrac{{\b}^2 J^2}{4} (P-1)\theta (\q_1^P-\q_0^P)
    \label{eq:ATSKK}
\end{align}
where 
\begin{align}
     g(\beta', J, \q_1, \q_0) =\b J z^{(1)} \sqrt{\dfrac{P}{2} \q_0^{P-1}} + \b J z^{(2)} \sqrt{\dfrac{P}{2} (\q_1^{P-1}-\q_0^{P-1})},
     \label{eq:gPSK}
\end{align}
$\mathbb{E}_1$, $\mathbb{E}_2$ are the average w.r.t. the standard Gaussian variables $z^{(1)}$ and $z^{(2)}$ and $\q_0, \ \q_1$ fulfill the following self-consistency equations 
\begin{equation}
\begin{array}{lll}
    \q_1 &= \mathbb{E}_1 \left[ \dfrac{\mathbb{E}_2 \cosh^\theta g(\beta', J, \q_1, \q_0) \tanh^2 g(\beta', J, \q_1, \q_0)}{\mathbb{E}_2 \cosh^\theta g(\beta', J, \q_1, \q_0)}\right], \\\\
    \q_0 &= \mathbb{E}_1 \left[ \dfrac{\mathbb{E}_2 \cosh^\theta g(\beta', J, \q_1, \q_0) \tanh (\beta', J,\q_1, \q_0)}{\mathbb{E}_2 \cosh^\theta g(\beta', J,\q_1, \q_0)}\right]^2.
\end{array}
\label{eq:self_RSB}
\end{equation}
From now on, we imply the dependence of $g$ and of the RS and 1RSB quenched free energies $f_{RS}$, $f_{1RSB}$ on $\beta'$ and $J$. The aim is to prove that the 1RSB approximation of the quenched free-energy is smaller than the replica symmetric one, above a critical value of the parameter $\beta' J$.

% , namely $f_{1RSB}(\beta, \theta) < f_{RS}(\beta)$ if
% \linda{\begin{equation}
% \label{eq:ATPspin}
% \begin{array}{lll}
%      &\dfrac{{\b}^2 (P-1) P }{2} \qq^{P-2} \mathbb{E}_1 \left\{\mathbb{E}_2\left[ \dfrac{1}{\cosh^2 g(\q, \qq)}\right]\right\}^2 > 1.
%      \end{array}
% \end{equation} 
%  In the limit $\qq \to \q$, this is the corresponding of the AT line in the SK model \cite{bardina2004}.}

As before, we note that $\lim_{\theta\to 1}\q_0=\q$ and 
$f_{1RSB} (\q_1, \q_0 \vert \theta=1)= f_{RS} (\q)$ and we aim at expanding the 1RSB quenched free-energy for $\theta\simeq 1$. 
% : 
% \begin{align}
%     -\b f_{1RSB} (\q_1, \q_0 {\vert \beta', J, \theta}) =&  - \b f_{RS} (\q \vert \b, J) + (\theta-1) \dfrac{\partial (-\b f_{1RSB})}{\partial \theta} \vert_{\theta=1}.
%     \label{eq:fexp}
% \end{align}
To this purpose, we first expand the self-consistency equations for $\q_0$ and $\q_1$ to linear order in $\theta-1$, to obtain 
\begin{align}
    \label{eq:q0exp}
    \q_0 &= \q + (\theta-1) A( \q, \qqq( \q))\\
    \q_1 &= \qqq( \q) + (\theta-1) B( \q, \qqq( \q))
    \label{eq:q1exp}
\end{align}
where $\qqq( \q)$ solves the self-consistency equation 
\begin{align}
    \q_1 = \mathbb{E}_1 \left\{ \dfrac{\mathbb{E}_2 \cosh g(\q_1, \q) \tanh^2 g(\q_1, \q) }{\mathbb{E}_2 \cosh g(\q_1, \q) }\right\}
    \label{eq:qqq_q}
\end{align} 
and $A(\q_0, \q_1)$ and $ B(\q_0, \q_1) $ are given \eqref{eq:A_Pspin_t1} and \eqref{eq:B_Pspin_t1}, respectively.
Then, we derive w.r.t. $\theta$ the 1RSB free-energy \eqref{eq:ATSKK} where we replace $\q_0$ and $\q_1$ with \eqref{eq:q0exp}, \eqref{eq:q1exp}, obtaining
\begin{align}
    \partial_\theta(-\b f_{1RSB}(\q_1, \q_0, \vert \theta)) &= -\dfrac{{\b}^2 J^2}{4} (P-1)((\tilde{q}_1(\q))^P - \q^P) - \dfrac{1}{\theta^2} \mathbb{E}_1 \ln \mathbb{E}_2 \cosh^\theta g(\qqq( \q), \q) \notag \\
    & + \dfrac{1}{\theta} \mathbb{E}_1 \left[ \dfrac{\mathbb{E}_2 \cosh^\theta g(\qqq( \q), \q)  \log \cosh g(\qqq(\q), \q)  }{\mathbb{E}_2 \cosh^\theta g(\qqq( \q), \q)  }\right] \nonumber \\
    &+ \dfrac{{\b}^2 J^2}{4} P (P-1) B(\qqq(\q), \q)  \theta \q^{P-2} \mathbb{E}_1 \left[ \dfrac{\mathbb{E}_2 \cosh^\theta g(\qqq( \q), \q)  \tanh g(\qqq( \q), \q)  }{\mathbb{E}_2 \cosh^\theta g(\qqq( \q), \q) }\right]^2 \nonumber \\
    & - \dfrac{{\b}^2 J^2}{4} P (P-1) B(\qqq( \q), \q)  \q^{P-1},
\end{align}
which, for $\theta=1$, using \eqref{eq:self_RS_PSpinHOP} and performing similar calculations to those in \eqref{eq:q-lim-theta1}, evaluates to
\begin{align}
    K(\tilde{q}_1(\q), \q):=  \partial_\theta(-\b f_{1RSB}(\q_1, \q_0, \vert \theta)) &\vert_{\theta=1} =-\dfrac{{\b}^2 J^2}{4} (P-1)((\tilde{q}_1(\q))^P - \q^P)  \notag \\
    & + \mathbb{E}_1 \left[ \dfrac{\mathbb{E}_2 \cosh g(\qqq( \q), \q)  \log \cosh g(\qqq( \q), \q)  }{\mathbb{E}_2 \cosh g(\qqq(\q), \q)  }\right] \nonumber \\
    &- \mathbb{E} \ln \cosh \left( \b \sqrt{\dfrac{P}{2} \q^{P-1}}\right)-\dfrac{{\b}^2 J^2}{4} P((\tilde{q}_1(\q))^{P-1} - \q^{P-1}).
    \label{eq:K_PSK}
\end{align}
%We stress that, for $\tilde{q}_1(\q) \to \q$, $K(\q, \q)=0$. 
Next, we study the sign of \eqref{eq:K_PSK}, where $\q$ and $\qqq( \q)$ are the solutions of the self-consistency equations \eqref{eq:self_RS_PSpinHOP} and \eqref{eq:qqq_q}, respectively. To this purpose, we study the behaviour of the function $K(x,\q)$ for $x \in [0, \q]$. For $x=\q$, we have $K(\q, \q)=0$, while the extremum of $K(\q, x)$ is found from
\begin{align}
    \partial_x K(x, \q) 
    % =& -\dfrac{{\b}^2 J^2}{4} (P-1) P x^{P-1}-\dfrac{{\b}^2 J^2}{4} (P-1)Px^{P-2} \notag \\
    % &+\dfrac{\b J P(P-1)x^{P-2}}{4\sqrt{P/2\  x^{P-1}}} \mathbb{E}_1\left[ \dfrac{\mathbb{E}_2\left( (\sinh g \log \cosh g + \sinh g )z^{(2)}\right)}{\mathbb{E}_2 \cosh g }\right] \notag \\
    % &- \dfrac{{\b}^2 J^2}{4} (P-1) P x^{P-2} \mathbb{E}_1 \left[ \dfrac{\mathbb{E}_2 \cosh g \log \cosh g }{\mathbb{E}_2 \cosh g }\right] \notag \\
    &=-\dfrac{{\b}^2 J^2}{4} (P-1) P x^{P-1} + \dfrac{{\b}^2 J^2}{4} (P-1) P x^{P-2} \mathbb{E}_1 \left[ \dfrac{\mathbb{E}_2 \sinh g(x, \q)  \tanh g(x, \q)  }{\mathbb{E}_2 \cosh g(x, \q)  }\right] =0,
\end{align}
%Therefore, we state that
as 
\begin{align}
   %\partial_x K(x, \q) = 0 \Leftrightarrow
   x= \mathbb{E}_1 \left[ \dfrac{\mathbb{E}_2 \sinh g(x, \q)  \tanh g(x, \q)  }{\mathbb{E}_2 \cosh g(x, \q)  }\right] \equiv \tilde{q}_1(\q)
\end{align}
from Eq. \eqref{eq:qqq_q}. Given that $K(x,\q)$ vanishes for $x=\q$, if the extremum $x=\qqq( \q)$ is global in the domain considered, we must have that $K(\tilde q_1(\q),\q)>0$ if $x=\tilde q_1(\q)$ is a maximum and $K(\qqq( \q), \q) <0$ if $x=\tilde q_1(\q)$ is a minimum. Therefore, if 
% The second derivative w.r.t. $x$, instead, is
% \begin{align}
%     \partial_{x^2} K(x, \q) =& - \dfrac{{\b}^2 J^2}{4} (P-1)^2 P x^{P-2} + \dfrac{{\b}^2 J^2}{4} (P-1) (P-2) P \tilde{q}_1 x^{P-3} \notag \\
%     &+ 2 \left( \dfrac{{\b}^2 J^2}{4} (P-1) P \right)^2 x^{2P-4}  - 2 \left( \dfrac{{\b}^2 J^2}{4} (P-1) P \right)^2 x^{2P-4}  \mathbb{E}_1 \left[ \dfrac{\mathbb{E}_2 \sinh g \tanh^3 g }{\mathbb{E}_2 \cosh g }\right] \notag \\
%     &- 4 \left( \dfrac{{\b}^2 J^2}{4} (P-1) P \right)^2 x^{2P-4} \tilde{q}_1 , 
% \end{align}
% which means, when $x=\tilde{q}_1$, that
\begin{align}
    \partial_{x^2} K(x, \q)\vert_{x=\tilde{q}_1} =& -\dfrac{{\b}^2 J^2}{4} (P-1) P \tilde{q}_1^{P-2} \left( 1-\dfrac{{\b}^2 J^2}{2} (P-1) P (\tilde{q}_1(\q))^{P-2} \mathbb{E}_1 \left[ \dfrac{\mathbb{E}_2 \textnormal{sech}^3 g(\qqq( \q), \q)   }{\mathbb{E}_2 \cosh g(\qqq( \q), \q)  }\right]\right)
    \label{eq:ATPspin}
\end{align}
is positive, $K(\qqq( \q), \q) $ is negative and the RS theory becomes unstable. This occurs for 
\begin{align}
    1-\dfrac{{\b}^2 J^2}{2} (P-1) P (\tilde{q}_1(\q))^{P-2} \mathbb{E}_1 \left[ \dfrac{\mathbb{E}_2 \textnormal{sech}^3\  g(\qqq(\q), \q) }{\mathbb{E}_2 \cosh g(\qqq( \q), \q) }\right] < 0.
    \label{eq:ATt1}
\end{align}

In the limit $\qqq( \q) \to \q$ this recovers the AT line found in \cite{gardner1985spin}
\begin{align}
\dfrac{(P-1){\b}^2 J^2 P \q^{P-2}}{2} \mathbb{E} \left[\dfrac{1}{\cosh^4 \sqrt{{\b}^2 J^2 P \q^{P-1} z/2}}\right] >1.
\label{eq:AT-Pspin}
\end{align}
By numerically solving the self-consistency equations  \eqref{eq:self_RS_PSpinHOP}
and \eqref{eq:qqq_q}, \resub{we can check that}
the parameters $\q$ and $\qqq( \q)$, 
%we find that these show deviations that increase with $P$
\resub{are indeed equal at all temperatures}
see Fig. \ref{fig:diffq0qSK} (mid and right panels), \resub{hence the above limit is justified a posteriori}. 
%\resub{Indeed, the system is known to undergo a discontinuous phase transition in the order parameter for $P>2$.}

\subsection{P-spin spherical model}
In this section we consider the spherical $P$-spin model, introduced for the first time in \cite{Crisanti}. The Hamiltonian of the model is the same as in \eqref{eq:HamiltonianPspin}, however the $N$ spins $\si$ are now real variables, 
satisfying the so-called 'spherical' constraint $\sum_{i=1}^N \si^2=N$. 
As in previous cases, the order parameter is the two-replica overlap $q$ introduced in \eqref{eq:def_m}. In the thermodynamic limit, at inverse temperature $\beta$, the quenched free-energy under the RS assumption  \eqref{eq:P_RS_Hop1} is given by
\begin{align}
    -2\beta f_{RS}(\q \vert \beta) = \dfrac{\beta^2}{2}(1-\q^P) + \log (1-\q) + \dfrac{\q}{1-\q},
\end{align}
where $\q$ fulfills the self-consistency equation
\begin{align}
     \dfrac{\beta^2}{2} P \q^{P-1} = \dfrac{\q}{(1-\q)^2}
     \label{eq:selfSPHRS1}
\end{align}
For later convenience, it is useful to note that the temperature $T^\star$ at which $\q$ becomes non-zero within the RS theory is found by demanding that \eqref{eq:selfSPHRS1} allows non-zero solutions satisfying
    \begin{align}
     \dfrac{2}{P} T^2=\q^{P-2}(1-\q)^2
     \label{eq:selfSPHRS2}
\end{align}
Denoting the RHS with $f(\q)$ and noting that $f(0)=f(1)=0$, a simple graphical argument shows that 
a non-zero solution exists when the LHS is smaller than 
$f(\q)$ evaluated at its maximum point $q=\dfrac{P-2}{P}$, i.e. for 
\iffalse
. Therefore 
\begin{align}
    f'(q)=(P (1 - q)-2) (1 - q) q^{P-3} = 0 \Rightarrow q=0 \ or\  q=1\  or \ q=\dfrac{P-2}{P}.
\end{align}
Since $f\left(\dfrac{P-2}{P}\right)>0=f(0)=f(1)$ the maximum point is $q=\dfrac{P-2}{P}$ and \eqref{eq:selfSPHRS2} is satisfied giving 
\fi
$T<T^\star$ with $T^\star=\sqrt{2(P-2)^{P-2}/P^{P-1}}$ \cite{Crisanti, Cavagna}. 

% $\q=(P-2)/P$, giving $T<T_c$ with $T_c=\sqrt{2(P-2)^{P-2}/P^{P-1}}$ \cite{Crisanti, Cavagna}. 
%
Within the 1RSB assumption \eqref{eq:1RSBAss},
the quenched free-energy evaluates to 
\begin{align}
\label{eq:f1RSBSP}
    -2\beta f_{1RSB}(\q_1, \q_0 \vert \beta, \theta)=&\dfrac{\beta^2}{2}\left[1+(\theta-1)\q_1^P -\theta \q_0^P \right] \notag \\
    &+\dfrac{\theta-1}{\theta} \log \left( 1-\q_1 \right)+ \dfrac{\q_0}{1-\q_1 + \theta (\q_1-\q_0)} \notag \\
    &+ \dfrac{1}{\theta} \log \left( 1-\q_1 + \theta (\q_1-\q_0)\right), 
\end{align}
where $\q_1$ and $\q_0$ fulfill the  self-consistency equations 
\begin{align}
    \dfrac{\beta^2}{2} P \q_0^{P-1} &= \dfrac{\q_0}{(1-\q_1 + \theta (\q_1-\q_0))^2}
    \label{eq:self_SPS_0_1}
    \\
    \dfrac{\beta^2}{2} P \q_1^{P-1} &= \dfrac{\beta^2}{2} P \q_0^{P-1} + \dfrac{\q_1-\q_0}{(1-\q_1 + \theta (\q_1-\q_0))^2}.
    \label{eq:self_SPS_1_1}
\end{align}
From now on, we imply the dependence of $f_{RS}$ and $f_{1RSB}$ on $\beta$. We note that for $\theta=1$, \eqref{eq:self_SPS_0_1} 
becomes equal to \eqref{eq:selfSPHRS1}, hence $\q_0(\theta=1)=\q$ and we also have 
$f_{1RSB}(\bar{q}_1,\bar{q}_0 \vert \theta)|_{\theta=1}=f_{RS}(\q)$. 
%Following the reasoning in Sec. \ref{sec:Hopfield}, we then assume that at the onset of the RS instability, $\theta$ is close to one so we 
The aim is to expand 
the 1RSB quenched free-energy around $\theta=1$ to linear orders in $\theta-1$. To this purpose, 
\iffalse
\begin{align} 
   f_{1RSB} (\bar{q}_1,\bar{q}_0 \vert \theta)=& f_{1RSB}(\bar{q}_1,\bar{q}_0 \vert \theta)|_{\theta=1}
   + (\theta-1) \partial_\theta f_{1RSB}(\bar{q}_1,\bar{q}_0 \vert \theta)\vert_{\theta=1},
    \label{eq:expansionPSP}
\end{align}
We observe that in principle, for $\theta=0$, we would {\it also} have $f_{1RSB}(\bar{q}_1,\bar{q}_0 \vert \theta)|_{\theta=0}=f_{RS}(\q)$ and $\q_1(\theta=0)=\q$ (by summing 
\eqref{eq:self_SPS_0_1} and \eqref{eq:self_SPS_1_1}). However, as shown later, expanding around 
$\theta=0$ one finds a critical temperature for the onset of the RS instability which is equal to $T^\star$ and \resub{smaller} than the critical temperature $T_c$ found by expanding around $\theta=1$, hence the latter is the physical one.  As explained in Sec. \ref{sec:Hopfield}, in systems where 
$T_c>T^\star$, at the onset of RS instability $\q=0$, hence only the scenario $\theta=1$ and 
$q_0(\theta=1)=\q$ can apply. 
%
Anticipating that $T^\star$ is smaller than the temperature $T_c$ at which the RS solution becomes unstable (a fact that is well-known from the literature \cite{Crisanti-RSB, Cavagna} and that we prove below within our approach)
we will assume that at the onset of the RS instability $\q=0$. 
%
Expanding 
\fi
we expand 
the 1RSB self-consistency equations around $\theta=1$ to obtain 
\begin{align}
     \dfrac{\beta^2}{2} P \q_0^{P-1} &=  \dfrac{\q_0}{(1-\q_0)^2} + (\theta-1) A(\q_0, \q_1) 
     \label{eq:SPSq1-A}
\end{align}
where $A(\q_0, \q_1)$ is defined in \eqref{eq:A-SPS} and 
\begin{align}
     \dfrac{\beta^2}{2} P \q_1^{P-1} &= \dfrac{\q_0}{(1-\q_0)^2}+ \dfrac{\q_1 - \q_0}{(1-\q_0)(1-\q_1)} + (\theta-1) B(\q_0, \q_1)
     \label{eq:q1SGS}
\end{align}
where $B(\q_0, \q_1)$ is as in \eqref{eq:B-SPS}.
As noted above, if $\theta=1$, $\q_0(\theta=1)=\q$, while 
 $\q_1(\theta=1)$ fulfills the following equation
\begin{align}
    \dfrac{\beta^2}{2} P \q_1^{P-1} &=  \dfrac{\q_1}{(1-\q_1)}
    \label{eq:q1-SPSp-theta1}
\end{align}
We will see below that the RS instability occurs at a temperature $T_c>T^\star$, hence the only solution of \eqref{eq:self_SPS_0_1} at $T_c$ is $\q_0=0$. For $\theta=1$, this corresponds to the paramagnetic solution $\q=0$. For $\theta<1$, the solution $\q_0=0$
remains valid as in the absence of external field all the states must be orthogonal to each other, leading to a vanishing mutual overlap \cite{Cavagna}.
%
% \iffalse
% \begin{align}
%     \dfrac{\beta^2}{2} P \q_1^{P-1} &= \dfrac{\q}{(1-\q)^2}+ \dfrac{\q_1 - \q}{(1-\q)(1-\q_1)}
%     \label{eq:q1-SPSp-theta1}
% \end{align}
% \fi
As the 1RSB theory requires $\q_1>\q_0$, we are interested in the non-zero solution of \eqref{eq:q1-SPSp-theta1}, which we can denote with $\tilde{q}_1$ and can be found explicitly from 
\begin{align}
\q_1^{P-2} (1-\q_1)=\frac{2}{P}T^2
\end{align}
Denoting the LHS with $g(\q_1)$ and noting that $g(0)=g(1)=0$, and reasoning as for $T^\star$, a non-zero solution exists when the RHS is smaller than 
$g(\q_1)$ evaluated at its maximum point 
$\q_1^\star=(P-2)/(P-1)$, giving $T<\sqrt{P(P-2)^{P-2}/2(P-1)^{P-1}}$ \cite{Cavagna}. 
Next, we compute the derivative w.r.t. $\theta$ of $-2\beta f_{1RSB}$ at $\q_0=\q$ and $\q_1=\qqq$: 
\begin{align}
    K(\tilde{q}_1,\q) =& \partial_\theta \left( -2\beta f_{1RSB}(\q_1,\q_0 \vert \theta)\right)\vert_{\theta=1} = \log \left( \dfrac{1-\tilde{q}_1}{1-\q}\right) \notag \\
    &+ (\q-\tilde{q}_1) \dfrac{P(1-\q)(1-2\q)(1-\tilde{q}_1)-\q(1+\q)-\tilde{q}_1(1-3\q)}{P(1-\q)^3(1-\tilde{q}_1)}
    %\dfrac{-\q^2 -P(\q-1/2)(\q-1)(1-\tilde{q}_1^*) - \tilde{q}_1^*-\q(1-3\tilde{q}_1^*)}{P(1-\q)^3(1-\tilde{q}_1^*))}
    \label{eq:KSGS}
\end{align}
Substituting $\q=0$ and the value of $\tilde{q}_1=\q_1^\star$ this evaluates to 
\iffalse
\begin{align}
     K(0, \q_1^*) =& \log(1-\q_1^*) + \q_1^* + \dfrac{(\q_1^*)^2}{P(1-\tilde{q}_1^*)}.
\end{align}
which can be written explicitly as 
\fi
\begin{align}
    K(\q_1^*,0)=2-\dfrac{4}{P}-\log(P-1)
\end{align}
which is always negative for $P>2$, implying
\begin{align}
    f_{1RSB}(\tilde{q}_1^*,0 \vert \theta)=f_{RS}(0)+(1-\theta)\left(\dfrac{K(0, \tilde{q}_1^*)}{2\beta}\right) < f_{RS}(0).
\end{align}
This shows that at the temperature $T_c=\sqrt{P(P-2)^{P-2}/2(P-1)^{P-1}}$, where a non-zero overlap $\q_1$ first emerges, the RS theory becomes unstable, as known in the literature \cite{Crisanti, Cavagna}. It can be easily verified that $T_c>T^\star$ for all $P>2$. 
% %
% Again, we want to study the function $K(0, x)$, $x \in [0, 1]$ in order to find possible extrema. We have that $K(0, 0)=0$ and 
% \begin{align}
%     \partial_x K(0, x)\vert_{x=\tilde{q}_1(0)}=1-\dfrac{1}{1-\tilde{q}_1(0)} - \dfrac{\tilde{q}_1(0)(\tilde{q}_1(0)-2)}{(1-\tilde{q}_1(0))^2}=0,
% \end{align}
% which has non-null solution for $\tilde{q}_1(0)= \dfrac{P-2}{P-1}$. Moreover, this expression gives us a condition also in temperature: indeed, using \eqref{eq:q1SGS} with $\q=0$, we get that
% \begin{align}
%     \dfrac{\beta^2}{2} P \tilde{q}_1(0)^{P-1}(1-\tilde{q}_1(0))=\dfrac{P-2}{P-1},
% \end{align}
% namely 
% \begin{align}
%     T^2=\dfrac{2(P-2)}{P(P-1)}\dfrac{\tilde{q}_1(0)^{1-P}}{(1-\tilde{q}_1(0))}
% \end{align}
% Furthermore, studying the second-order derivative w.r.t. x in $\tilde{q}_1(0)= \dfrac{P-2}{P-1}$ we get 
% \begin{align}
%     \partial_x^2 K(\q, x)\vert_{x=\dfrac{P-2}{P-1}} = \dfrac{(P-2)(P-1)^2}{P}, 
% \end{align}
% which is always positive when $P>2$. Therefore, $K(0, \tilde{q}_1(0))>0$ but $f_{1RSB}>f_{RS}$. 

\section{Expanding around $\theta=0$}
\label{app:expt0}
Although we have so far regarded the limit 
$\theta\to 1$ (where $\q_0=\q$ and $f_{\rm 1RSB}=f_{\rm RS}$) as the physical one, in the opposite limit, 
$\theta\to 0$, we would {\it equally} find, for all the models considered above, $f_{\rm 1RSB}=f_{\rm RS}$ (with $\q_1=\q$), suggesting 
that a similar analysis could have been carried for $\theta\to 0$.

In this section we present such analysis for the Hopfield model, the Hebbian networks with multi-node interactions and the spherical $P$-spin. The same analysis can be carried out for the other spin-glass models considered in Appendix \ref{app:spin-glasses}. Given the strong similarity of the SK and the Ising $P$-spin models with the Hopfield and the dense associative memory models, respectively, we will not report such analysis here.

\subsection{The Hopfield model}
\label{sec:Hopfield-theta0}
From \eqref{eq:self_RSBHOP}, one finds
\begin{align}
    \lim_{\theta\to 0}\q_1&=  \lim_{\theta\to 0}\mathbb{E}_1 \mathbb{E}_2 \tanh^2 \left(\beta \m+
\beta z^{(1)} \frac{\sqrt{\alpha \q_0}}{\Delta_2(\theta, \q_0, \q_1)} \right.
+ \left.\beta z^{(2)} \sqrt{\alpha \frac{\q_1-\q_0}{\Delta_1(\q_1)  \Delta_2(\theta,\q_0, \q_1)}}
    \right)
    \nonumber\\
&=    \mathbb{E}_1 \mathbb{E}_2 \tanh^2 \left(
    \beta \m + \beta z^{(1)} \dfrac{\sqrt{\alpha \q_0}}{\Delta_1(\q_1)} + \beta z^{(2)} \dfrac{\sqrt{\alpha(\q_1-\q_0)}}{\Delta_1(\q_1)}
    \right)\nonumber\\
    &=\mathbb{E} \tanh^2 \left( \beta \m + \beta \sqrt{\dfrac{\alpha \q_1}{(1-\beta(1-\q_1))^2}} z\right)%\equiv\bar{q}
    \label{eq:q-lim-theta0}
\end{align}
where we have used that for $\theta=0$, $\Delta_2(0, \q_0, \q_1)=\Delta_1(\q_1)$ and the relation 
\begin{align}
\label{eq:relation}
    \mathbb{E}_{\lambda,Y}[F(a_1+ \lambda a_2+Y a_3)]=\mathbb{E}_{_Z}\left[F\left(a_1+Z\sqrt{a_2^2+a_3^2}\right)\right],
\end{align}
with $F$ any smooth function, $a_1, \ a_2,\ a_3 \in \mathbb{R}$, and $\lambda$, $Y$ and $Z$ i.i.d. standard normal random variables. As \eqref{eq:q-lim-theta0}
 is identical to \eqref{eq:AGSq}, in the limit $\theta\to 0$, $\q_1$ is equal to the RS order parameter $\q$. 
 \resub{Similarly, one can show that 
\begin{align}
    \lim_{\theta\to 0}\mm&= \m
    \end{align}
    }
 and can easily verify that $f_{1RSB}(\resub{\mm,\,}\bar{q}_{1},\bar{q}_0 \vert \theta )|_{\theta=0}=f_{RS}(\resub{\m,\,}\bar{q})$. %, as expected from the fact that, for $\theta=0$, eq. (\ref{eq:1RSBAss}) reduces to (\ref{eq:P_RS_Hop2}) and one retrieves the RS scheme.
Our purpose is then to prove that for small but finite values of $\theta$ the 1RSB expression of the quenched 
%statistical pressure 
free-energy is 
%smaller 
smaller
than the RS expression, i.e. 
$f_{1RSB}(\resub{\mm,\,}\bar{q}_{1},\bar{q}_0\vert \theta)<f_{RS}(\resub{\m,\,}\bar{q})$, 
below a critical line in the parameters space $(\alpha,\beta)$.

To this purpose, we expand the 1RSB quenched 
free-energy 
around $\theta=0$ -namely around the replica symmetric expression- to the first order, to write
\begin{align} 
   f_{1RSB} (\resub{\mm,\,}\bar{q}_1,\bar{q}_0 \vert \theta)=& f_{1RSB}(\resub{\mm,\,}\bar{q}_1,\bar{q}_0 \vert \theta)|_{\theta=0}
   + \theta \partial_\theta f_{1RSB}(\resub{\mm,\,}\bar{q}_1,\bar{q}_0 \vert \theta)\vert_{\theta=0},
    \label{eq:expansionHOP}
\end{align}
where $f_{1RSB}(\resub{\mm,\,}\bar{q}_1,\bar{q}_0\vert \theta)|_{\theta=0}= f_{RS}(\resub{\m,\,}\bar{q})$. To determine when the RS solution becomes unstable, i.e.  $f_{1RSB}(\resub{\mm,\,}\bar{q}_1,\bar{q}_0 \vert \theta)<f_{RS}(\resub{\m,\,}\bar{q})$ we inspect the sign of $\partial_\theta f_{1RSB}(\resub{\mm,\,}\bar{q}_1,\bar{q}_0 \vert \theta)\vert_{\theta=0}$.
To evaluate the latter, we need to expand the self-consistency equations for $\q_0$, $\q_1$ \resub{and $\mm$} around $\theta=0$ to linear orders in $\theta$. Using \eqref{eq:q-lim-theta0} and denoting 
\begin{align}
\label{eq:g0HOP}
g_0(\resub{\mm,\,}\q_1, \q_0) = \beta \mm + \beta z^{(1)} \dfrac{\sqrt{\alpha \q_0}}{\Delta_1(\q_1)} + \beta z^{(2)} \dfrac{\sqrt{\alpha(\q_1-\q_0)}}{\Delta_1(\q_1)},
\end{align}
we obtain
\begin{align}
\label{eq:q1}
    \q_1 &= \mathbb{E}_1 \mathbb{E}_2 \tanh^2 g_0(\resub{\m,\,}\q_1, \q_0) 
    + \theta A(\resub{\mm,\,}\q_0,\q_1) 
%    \notag \\%+\mathcal{O}(\theta^2) =  
%    &=: \mathbb{E} \tanh^2 \left( \beta \m + \beta J \sqrt{\dfrac{\alpha \q_1}{1-\beta(1-\q_1)}}z\right) + \theta A(\q_1, \q_0) 
    \end{align}
    where $A(\resub{\mm,\,}\q_0,\q_1)$ is a function of $\q_0$ and $\q_1$ that will drop out of the calculation, whose expression is provided in \eqref{eq:A-Hopfield}.
It follows from \eqref{eq:q-lim-theta0}
     that to $\mathcal{O}(\theta^0)$, 
    $\q_1$ is equal to the RS order parameter $\q$ so
    we can rewrite \eqref{eq:q1}
    %, applying \eqref{eq:relation}, 
    as 
    \begin{equation}
        \q_1= \q+\theta A(\resub{\mm,\,}\q_0, \q).
        \label{eq:q1-q-0}
    \end{equation}
    Following the same path for $\q_0$, and using \eqref{eq:q1-q-0}, we have 
    \begin{align}
    \q_0 
    %&= \mathbb{E}_1\left( \mathbb{E}_2 \tanh g(\q_0, \q) \right)^2 + \theta B(\q_1,\q_0)\notag \\
    % &=: \mathbb{E}_1\left( \mathbb{E}_2 \tanh g_0(\q_0, \q) \right)^2 + \theta B(\q_1, \q_0)
        &= \qq(\resub{\m,\,}\q)+ \theta B(\resub{\mm,\,}\q_0, \q)
    \label{eq:q0-q-0}
    \end{align}
where $B(\resub{\mm,\,}\q_0,\q)$ is provided in \eqref{eq:B-Hopfield} and will drop out of the calculation, and
we have denoted with $\tilde q_0(\resub{\m,\,}\q)$ the solution of 
\begin{align}
    \q_0=\mathbb{E}_1\left( \mathbb{E}_2 \tanh g_0(\resub{\m,\,}\q_0, \q) \right)^2. %= \qq
    \label{eq:self2-0}
\end{align}
Finally, we can write the magnetization as 
\begin{align}
    \mm = \m + \theta C(\resub{\m,\,} \q,\resub{\tilde q_0(\m,\q)})
    \label{eq:m-0}
\end{align}
where $C(\resub{\m,\,} \q,\resub{\tilde q_0(\m,\q)})$ is given in \eqref{eq:C-Hopfield} \resub{and rewrite \eqref{eq:q1-q-0} and \eqref{eq:q0-q-0} as 
    \begin{eqnarray}        
    \q_1&=& \q+\theta A(\resub{\m,\,}\tilde q_0(\m,\q), \q)
        \label{eq:q1-q-0-fin}
        \\
    \q_0 &=&\qq(\resub{\m,\,}\q)+ \theta B(\resub{\m,\,}\tilde q_0(\m,\q), \q)
    \label{eq:q0-q-0-fin}
    \end{eqnarray}}
Using \eqref{eq:q1-q-0-fin}, \eqref{eq:m-0} and \eqref{eq:q0-q-0-fin} to  evaluate the derivative of 
$f_{1RSB}(\resub{\mm,\,} \bar{q}_1,\q_0 \vert \theta)$
w.r.t. $\theta$ and finally setting  $\theta=0$, we obtain:
\begin{align}
    &K(\resub{\m,\,} \q, \qq(\m,\q)):=\partial_\theta (-\beta f_{1RSB}(\resub{\mm,\,} \q_1, \q_0 \vert \theta) )\vert_{\theta=0} 
    \notag\\
    &=  - \dfrac{\alpha \beta^2 (\q^2\!-\!{\qq(\resub{\m,\,}\q)}^2)}{4\Delta_1(\q)^2} 
    \!+\! \dfrac{1}{2} \mathbb{E}_1 \mathbb{E}_2 \ln^2 \cosh  g_0(\resub{\m,\,}\q, \qq(\resub{\m,\,}\q)) 
    - \dfrac{1}{2} \mathbb{E}_1 \left( \mathbb{E}_2 \ln \cosh g_0(\resub{\m,\,}\q, \qq(\resub{\m,\,}\q)) \right)^2
    \label{eq:K-0}
\end{align}
Next, we study the sign of \eqref{eq:K-0}, where $\q$ and $\qq(\resub{\m,\,}\q)$ are the solutions of the self-consistency equations \eqref{eq:AGSq} and \eqref{eq:self2-0}, respectively. To this purpose, it is useful to study the behaviour of the function $K(\resub{\m,\,}\q, x)$ for $x \in [0, \q]$. For $x=\q$, we have $K(\resub{\m,\,} \q, \q)=0$, \resub{regardless of the value assigned to $\m$}, 
while the extremum of $K(\resub{\m,\,}\q,x)$ is found from 
\begin{align}
    \partial_{x} K(\resub{\m,\,}\q, x) &= \dfrac{\beta^2 \alpha x}{2\Delta_1(\q)^2 } \left[ x- \mathbb{E}_1 \left( \mathbb{E}_2 \tanh g_0(\resub{\m,\,}\q, x) \right)^2 \right]=0
 \notag   \end{align}
    as 
    \begin{align}
x= \mathbb{E}_1 \left( \mathbb{E}_2 \tanh g_0(\resub{\m,\,}\q, x)\right)^2 \equiv \qq(\resub{\m,\,}\q), 
\end{align}
from Eq. \eqref{eq:self2-0}. Given that $K(\resub{\m,\,}\q,x)$ vanishes for $x=\q$, if the extremum $x=\qq(\resub{\m,\,}\q)$ is global in the domain considered, we must have that $K(\resub{\m,\,}\bar q, \tilde q_0(\resub{\m,\,}\q))>0$ if $x=\tilde q_0(\resub{\m,\,}\q)$ is a maximum and $K(\resub{\m,\,}\bar q, \tilde q_0(\resub{\m,\,}\q))<0$ if $x=\tilde q_0(\resub{\m,\,}\q)$ is a minimum. Therefore, if 
\begin{align}
    \partial_{x}^2 K(\resub{\m,\,}\q, x) \vert_{x=\qq(\resub{\m,\,}\q)} & = 
    \dfrac{\beta^2 \alpha }{2\Delta_1(\q)^2} \left\{ 1-\dfrac{\beta^2 \alpha}{\Delta_1(\q)^2} \mathbb{E}_1 \left\{ \mathbb{E}_2 \left[ \dfrac{1}{\cosh^2  g_0 (\resub{\m,\,}\qq(\resub{\m,\,}\q), \q)}\right]\right\}^2\right\}
    \label{eq:second-der-0}
\end{align}
is negative, $K(\resub{\m,\,}\q, \qq(\resub{\m,\,}\q))$ is positive and $f_{1RSB}(\resub{\m,\,}\q,\qq(\resub{\m,\,}\q)\vert \theta) < f_{RS} (\resub{\m,\,}\q)$,
hence the RS theory becomes unstable
when the expression in the curly brackets in \eqref{eq:second-der-0}
becomes negative i.e. for
\begin{equation}
     (1-\beta(1-\q))^2 < {\beta^2 \alpha} \mathbb{E}_1 \left\{ \mathbb{E}_2\left[ \dfrac{1}{\cosh^2  g_0(\resub{\m,\,}\q,\qq(\resub{\m,\,}\q))}\right]\right\}^2
     \label{eq:AT-hop-0}
\end{equation}
Interestingly, also in this case, the result found by Coolen in \cite{coolen2001statistical} using the de Almeida and Thouless' approach \cite{de1978stability}, is recovered from the expression above in the 
limit $\qq(\resub{\m,\,}\q)\to \q$. 
Solving numerically 
$\q$ and $\qq(\resub{\m,\,}\q)$ from the self-consistency equations \eqref{eq:AGSq} and \eqref{eq:self2-0}, respectively, one can verify  
that these two quantities are \resub{indeed identical} for any temperature, and the resulting RS instability line coincides with the classical AT line and the critical line given in \eqref{eq:AT-hop_t1}, obtained by expanding around $\theta=1$, see Fig. \ref{fig:exp01} (left panel). \resub{We anticipate that this will remain the case for Hebbian networks with $P$-node interactions, that we will analyse in the next section (see mid and right panels of Fig. \ref{fig:exp01})}. Although we do not report such analysis here, we have 
checked that this is also the case for the SK model.

\begin{figure}[t]
    \centering
    \includegraphics[width=15cm]{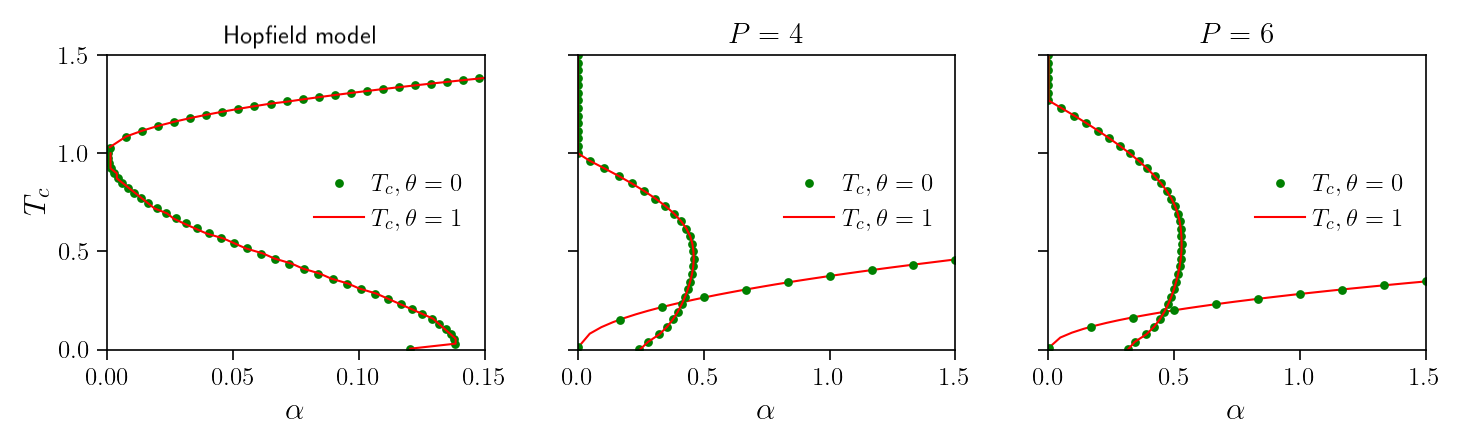}
    \caption{RS instability lines for the Hopfield model (left) and Hebbian networks with $P$-node interactions, with $P=4$ (mid) and $P=6$ (right), obtained via our method for $\theta \to 0$ and $\theta \to 1$. In the all the case the lines obtained for $\theta\to 0$ and $\theta\to 1$ are indistinguishable. 
    }
    \label{fig:exp01}
\end{figure}

\subsection{Hebbian Networks with multi-node interactions}
\label{app:expt0_Dense}

Here we apply the same analysis to Hebbian networks with multi-node interactions, defined by the Hamiltonian given in Eq. \eqref{eq:H_Dense}. Our objective is to prove that the 1RSB quenched free-energy is smaller than its replica symmetric counterpart i.e. $f_{1RSB}(\b, \alpha, \theta) < f_{RS}(\b, \alpha)$ above a critical value of the effective parameter $\sqrt{\alpha} \beta'$.
To this purpose we expand, to linear orders in $\theta$, the 1RSB quenched free-energy around $\theta=0$, as shown in \eqref{eq:expansionHOP}.
Since the self-consistency equations also depend on $\theta$, we need to expand them too. 
Following the same steps as in the Hopfield model, we can write \resub{$\mm$ as in \eqref{eq:m-0}, with $C(\resub{\m,\,} \q,\tilde q_0(\m,\q))$ as in \eqref{eq:C-Dense},} $\q_1$ as in \eqref{eq:q1-q-0}, with $A(\resub{\m,\,}\q,\qq(\resub{\m,\,}\q))$ given in \eqref{eq:A-Dense}, and $\q_0$ as given in \eqref{eq:q0-q-0}, 
where $\qq(\resub{\m,\,}\q)$ is the solution of the self-consistency equation %\eqref{eq:self2-0}
\begin{align}
     \q_0&=\mathbb{E}_1\left( \mathbb{E}_2 \tanh g(\resub{\m,\,}\q, \q_0) \right)^2 
     \label{eq:q0Dense}
\end{align}
and  $B(\resub{\m,\,}\q,\qq(\resub{\m,\,}\q))$ is given in \eqref{eq:B-Dense}.
With the above expressions in hand, we can now calculate the derivative of $f_{\rm 1RSB}$ w.r.t. $\theta$ when $\theta=0$, as needed in \eqref{eq:expansionHOP}
\begin{align}
\label{eq:KDense}
    &K(\resub{\m,\,}\q, \qq(\resub{\m,\,}\q)):=\partial_\theta (-\b f_{\rm 1RSB}(\resub{\mm,\,}\q_1, \q_0\vert \theta) )\vert_{\theta=0} 
    \notag\\
    &= \dfrac{-{\alpha\b}^2(P\!-\!1)}{4}  (\q^P \!-\!({\qq}(\resub{\m,\,}\q))^P) \!+\! \dfrac{1}{2} \mathbb{E}_1 \mathbb{E}_2 \ln^2 \cosh  g(\resub{\m,\,}\q, \qq(\resub{\m,\,}\q)) - \dfrac{1}{2} \mathbb{E}_1 \left( \mathbb{E}_2 \ln \cosh  g(\resub{\m,\,}\q, \qq(\resub{\m,\,}\q))  \right)^2
\end{align}
Again, we have that $K(\resub{\m,\,}\q, \q)=0$, \resub{regardless of the value assigned to $\m$} 
 (this follows from the fact that for $\theta=0$, $\q$ is an extremum of the free-energy). Next, we inspect the sign of $K(\resub{\m,\,}\q, \qq(\resub{\m,\,}\q))$. To this purpose, we study $K(\resub{\m,\,}\q, x)$ for $x \in [0, \q]$ and locate its extrema, which are found from 
\begin{align}
    \partial_{x} K(\resub{\m,\,}\q, x)=& \dfrac{{\b}^2 \alpha P(P-1)}{4} {x}^{P-2} \left[ x - \mathbb{E}_1\left( \mathbb{E}_2 \tanh  g (\resub{\m,\,}\q, x)\right)^2\right] =0
    \end{align}
    as 
\begin{align} x = \mathbb{E}_1\left( \mathbb{E}_2 \tanh  g(\resub{\m,\,}\q, x) \right)^2 \equiv \qq(\resub{\m,\,}\q)
\end{align}
where the last equality follows from \eqref{eq:q0Dense}. 
Under the assumption that 
the extremum $x=\qq(\resub{\m,\,}\q)$ is global in the domain considered and reasoning as in the 
Hopfield case, we have that 
$K(\resub{\m,\,}\bar q, \qq(\resub{\m,\,}\q))>0$ if $x=\tilde q_0(\resub{\m,\,}\q)$ is a maximum and $K(\resub{\m,\,}\bar q, \qq(\resub{\m,\,}\q))<0$
if it is a minimum. In particular, if 
\begin{align}
    \partial^2_{x} K(\resub{\m,\,}\q,x)\vert_{x=\qq(\resub{\m,\,}\q)}=& - \dfrac{{\b}^2 \alpha P(P-1)}{4} {\qq(\resub{\m,\,}\q)}^{P-2}\notag \\
    &\cdot\left\{ 1- \dfrac{{\b}^2 \alpha P(P-1) {(\qq(\resub{\m,\,} \q))}^{P-2}}{2} \mathbb{E}_1 \left[ \mathbb{E}_2 \dfrac{1}{\cosh^2 g(\resub{\m,\,}\q, \qq(\resub{\m,\,}\q))}\right]^2\right\}
    \label{eq:K2-Dense1}
\end{align}
is negative, $K(\resub{\m,\,}\q,\qq(\resub{\m,\,}\q))>0$ and $f_{1RSB}<f_{RS}$. This happens when the expression in the curly brackets of the equation above is negative, i.e. when the parameter $\alpha {\b}^2$ satisfies the inequality
\begin{equation}\label{result_PHOP}
\dfrac{\alpha{\b}^2 P(P-1) (\qq(\resub{\m,\,}\q))^{P-2}  }{2}\mathbb{E}_1\left\{ \mathbb{E}_2 \left[ \dfrac{1}{\cosh^2 g(\resub{\m,\,}\q, \qq(\resub{\m,\,}\q))}\right]\right\}^2 > 1.
\end{equation}
%Also in this case, in the limit $\qq(\q)\to \q$ the above expression retrieves the classical AT line \eqref{eq:AT-Pspin} found in \cite{gardner1985spin}, however by numerically solving the self-consistency equations, differences can be appreciated between $\q$ and $\qq(\q)$, hence this limit cannot be justified. 
The resulting critical line  
\resub{is found to be identical to}
the 
critical line \eqref{eq:result_PHOP} obtained from the 
expansion around $\theta=1$ (see Fig. \ref{fig:exp01}, mid and right panels).

\subsection{Spherical $P$-spin}
%We conclude this section by showing that the expansion around $\theta=0$ would provide, for the RS instability, a lower temperature than $T_c$, equal to $T^\star$, from which it can be concluded that the RS instability occurs at $T_c$.
Expanding for small $\theta$ the self-consistency equations \eqref{eq:self_SPS_0_1}, \eqref{eq:self_SPS_1_1} to linear orders, we get 
\begin{align}
    \dfrac{\beta^2}{2} P \q_1^{P-1} &= \dfrac{\beta^2}{2} P \q_0^{P-1} + \dfrac{\q_1-\q_0}{(1-\q_1 )^2} 
    + \theta A(\q_0, \q_1)
    \label{eq:SPS-theta0-1}
    \\
    \dfrac{\beta^2}{2} P \q_0^{P-1} &= \dfrac{\q_0}{(1-\q_1 )^2}+ \theta B(\q_0, \q_1)
    \label{eq:SPS-theta0-0}
\end{align}
%
\iffalse
\begin{align}
     \dfrac{\beta^2}{2} P \q_1^{P-1} 
     &= \dfrac{\q_1}{(1-\q_1)^2} + \theta A(\q_0, \q_1),
\end{align}
\fi
where the expression for $A(\q_0, \q_1)$ and $B(\q_0,\q_1)$ are provided in 
\eqref{eq:A-SPS0} and \eqref{eq:B-SPS0}, respectively. 
If $\theta =0$, summing the two equations gives 
\begin{align}
     \dfrac{\beta^2}{2} P \q_1^{P-1} 
     &= \dfrac{\q_1}{(1-\q_1)^2}
         \label{eq:self_SPSGq1}
\end{align}
showing that, to orders ${\mathcal O}(\theta^0)$,  $\q_1=\q$, while  
$\q_0$ fulfills the following self-consistency equation
\begin{align}
    \dfrac{\beta^2}{2} P \q_0^{P-1} &= \dfrac{\q_0}{(1-\q_1)^2}
    \label{eq:self_SPSGq0}
\end{align}
whose solution is denoted with $\qq(\q)$. 
The latter equation is solved by $\qq(\q)=0$ (which corresponds to the paramagnetic solution and remains valid when ergodicity is broken, as explained earlier). 
Similarly, $\q_1=0$ is always a solution of \eqref{eq:self_SPSGq1}, 
however the 1RSB scenario requires $\q_1>0$. As explained in the previous section, such non-zero solution appears at $T\leq T^\star=\sqrt{2(P-2)^{P-2}/P^{P-1}}$, which is below $T_c$ for any $P>2$, hence we can immediately conclude that the instability of the RS theory occurs at the larger temperature $T_c$, without further comparing the free-energies $1$RSB and RS for $\theta$ close to zero. 

\section{Contributions to sub-leading orders}
\label{app:subleading}
In this appendix we provide expressions for all the functions that we left unspecified in the main text, as they did not contribute to leading orders, 
including the  
functions $A(\q_0,\q_1)$ and $B(\q_0,\q_1)$ for all the models considered. 
\begin{itemize}
\item For the Hopfield model, in the expansion around $\theta=1$ the subleading contributions to \eqref{eq:q0_t1}
%\eqref{eq:q1-q_t1} 
and \eqref{eq:expanded-self_t1}
%\eqref{eq:q0-q_t1} 
are, \begin{align}
A(\resub{\m,\,} \q_0,\q_1)= 2
\mathbb{E}_1 &\left[ \tanh \left(  \beta \m + \dfrac{\beta z^{(1)}\sqrt{\alpha \q_0}}{(1-\beta(1-\q_0)}\right) \right.\notag \\
&\left.\cdot \dfrac{\mathbb{E}_2 \log \cosh g_1(\resub{\m,\,}\q_0, \q_1) \sinh g_1(\resub{\m,\,}\q_0, \q_1) (1- \tanh g_1(\resub{\m,\,}\q_0, \q_1)) }{\mathbb{E}_2 \cosh g_1(\resub{\m,\,}\q_0, \q_1)}\right],
    \label{eq:A-Hopfield_t1}
    \end{align}
        \begin{align}
&B(\resub{\m,\,} \q_0,\q_1)=  \dfrac{\beta^3 \alpha (\q_1-\q_0)}{(1-\beta(1-\q_0))^2} \left( \dfrac{\q_0}{(1-\beta(1-\q_0))} + \dfrac{(\q_1-\q_0)}{(1-\beta(1-\q_1))}\right)  \notag \\
&\cdot\mathbb{E}_1 \left\{ 1+ \dfrac{\mathbb{E}_2 \tanh^2 g_1(\resub{\m,\,}\q_0, \q_1)}{\mathbb{E}_2 \cosh g_1(\resub{\m,\,}\q_0, \q_1)} + \dfrac{\mathbb{E}_2 \log \cosh g_1(\resub{\m,\,}\q_0, \q_1) (1-\tanh^2 g_1(\resub{\m,\,}\q_0, \q_1))}{\mathbb{E}_2 \cosh g_1(\resub{\m,\,}\q_0, \q_1)}\right\}  \notag \\
&- \dfrac{\beta^3 \alpha (\q_1-\q_0)^2}{(1-\beta(1-\q))^3} \mathbb{E}_1 \left\{ \tanh^2 g_1(\resub{\m,\,}\q_0, \q_1)\left( 1- \right. \right. \left. \dfrac{\mathbb{E}_2 \sinh g_1(\resub{\m,\,}\q_0, \q_1) \tanh^2 g_1(\resub{\m,\,}\q_0, \q_1)}{\mathbb{E}_2 \cosh g_1(\resub{\m,\,}\q_0, \q_1)}\right)\notag \\
&+ \left(2-3 \tanh^3 \left(  \beta \m + \dfrac{\beta z^{(1)}\sqrt{\alpha \q_0}}{(1-\beta(1-\q_0)}\right)\right) \dfrac{\mathbb{E}_2 \sinh g_1(\resub{\m,\,}\q_0, \q_1) \tanh^2  g_1(\resub{\m,\,}\q_0, \q_1)}{\mathbb{E}_2 \cosh  g_1(\resub{\m,\,}\q_0, \q_1)} \notag \\
&\left.+ \tanh \left(  \beta \m + \dfrac{\beta z^{(1)}\sqrt{\alpha \q_0}}{(1-\beta(1-\q_0)}\right) \dfrac{\mathbb{E}_2 \log \cosh  g_1(\resub{\m,\,}\q_0, \q_1)\tanh  g_1(\resub{\m,\,}\q_0, \q_1)}{\mathbb{E}_2 \cosh  g_1(\resub{\m,\,}\q_0, \q_1)}\right\} \notag \\
&\dfrac{\beta^3 \alpha(\q_1-\q_0)^2}{2(1-\beta(1-\q_0))^2(1-\beta(1-\q_1))} \mathbb{E}_1 \left\{ \tanh \left( \beta \m +  \dfrac{\beta z^{(1)}\sqrt{\alpha \q_0}}{(1-\beta(1-\q_0)}\right)\right. \notag \\
&\left.\cdot \dfrac{\mathbb{E}_2 \sinh g_1(\resub{\m,\,}\q_0, \q_1) \tanh  g_1(\resub{\m,\,}\q_0, \q_1)}{\mathbb{E}_2 \cosh  g_1(\resub{\m,\,}\q_0, \q_1)}\right\} \notag \\
&- \mathbb{E}_1 \left\{ \dfrac{\mathbb{E}_2 \sinh g_1(\resub{\m,\,}\q_0, \q_1) \tanh  g_1(\resub{\m,\,}\q_0, \q_1) \mathbb{E}_2 \cosh  g_1(\resub{\m,\,}\q_0, \q_1) \log \cosh  g_1(\resub{\m,\,}\q_0, \q_1)}{(\mathbb{E}_2 \cosh  g_1(\resub{\m,\,}\q_0, \q_1))^2}\right\},
    \label{eq:B-Hopfield_t1}
\end{align}
and
\begin{align}
    &C(\resub{\m,\,} \q,\q_1)=\left(-3 \beta \dfrac{\sqrt{\alpha \q}}{\Delta_1(\q)} - \beta \sqrt{\dfrac{\alpha(\q_1-\q)}{\Delta_1(\q)\Delta_1(\q_1)}} \right)\mathbb{E}\tanh\left( \beta \m + \beta \sqrt{\dfrac{\alpha \q}{(1-\beta(1-\q))^2}} z \right) 
    \notag \\
    &+ 2 \beta \mathbb{E}\tanh^3\left( \beta \m + \beta \sqrt{\dfrac{\alpha \q}{(1-\beta(1-\q))^2}} z \right)\dfrac{\sqrt{\alpha \q}}{\Delta_1(\q)} \notag \\
    &- \beta \dfrac{\sqrt{\alpha \q}}{\Delta_1(\q)} \mathbb{E}_1 \left[ \sinh g_1(\m, \q_0, \q_1)\dfrac{\mathbb{E}_2 \cosh g_1(\m, \q_0, \q_1)\tanh g_1(\m, \q_0, \q_1) \log \cosh g_1(\m, \q_0, \q_1)}{\mathbb{E}_2 \cosh g_1(\m, \q_0, \q_1)}\right] \notag \\
    &+ \left( \beta \dfrac{\sqrt{\alpha \q}}{\Delta_1(\q)}+\beta \sqrt{\dfrac{\alpha(\q_1-\q)}{\Delta_1(\q)\Delta_1(\q_1)}}\right) \left(\mathbb{E}_1 \left[ \dfrac{\mathbb{E}_2 \cosh g_1(\m, \q_0, \q_1)\log \cosh g_1(\m, \q_0, \q_1) }{\mathbb{E}_2 \cosh g_1(\m, \q_0, \q_1)}\right]\right. \notag \\
    &\left.+ \mathbb{E}_1 \left[ \dfrac{2 \cosh g_1(\m, \q_0, \q_1) \tanh^2 g_1(\m, \q_0, \q_1)}{\mathbb{E}_2 \cosh g_1(\m, \q_0, \q_1)}\right]\right)
    \label{eq:C-Hopfield_t1}
\end{align}
respectively, 
where $g_1(\resub{\m,\,}\q_0, \q_1)$ is defined as in \eqref{eq:g1HOP}. 

For the expansion around $\theta=0$ the subleading contributions to \eqref{eq:q1} and 
\eqref{eq:q0-q-0} are
\begin{align}
A(\resub{\m,\,}\q_0,\q_1)=&
\mathbb{E}_1 \left[\mathbb{E}_2 \tanh^2 g_0(\resub{\m,\,}\q_1, \q_0) \ln \cosh g_0(\resub{\m,\,}\q_1, \q_0)\right] \notag \\
&- \mathbb{E}_1 \left[ \mathbb{E}_2 \tanh^2 \mathbb{E}_2 \ln \cosh g_0(\resub{\m,\,}\q_1, \q_0)\right]
    \notag\\
    &+\dfrac{\beta^3 \alpha \q_1(\q_1-\q_0)}{\Delta_1(\q_1)^3} \left[ 1-\mathbb{E}_1 \left( \mathbb{E}_2 \tanh^2 g_0(\resub{\m,\,}\q_1, \q_0)\right)\right] \notag \\
    &+ \dfrac{3\beta^3 \alpha (\q_1^2-\q_0^2)}{\Delta_1(\q_1)^3} \mathbb{E}_1 \left[ \mathbb{E}_2 \tanh^2 g_0(\resub{\m,\,}\q_1, \q_2) \left( 1-\tanh^2 g_0(\resub{\m,\,}\q_1, \q_0)\right)\right]
    % &+ \dfrac{\beta^3 (\q_1^2-\q_0^2)}{\Delta_1(\qq, \q) ^3} \mathbb{E}_1 \mathbb{E}_2 \left[ (1-3\tanh^2 g_0(\q_1, \q_0))
    % (1-\tanh^2 g_0(\q_1, \q_0))\right],
    \label{eq:A-Hopfield}
    \end{align}
    \begin{align}
&B(\resub{\m,\,}\q_0,\q_1)=  2  \Big\{ \mathbb{E}_1 \left[ \mathbb{E}_2 \tanh g_0(\resub{\m,\,}\q_1, \q_0) \mathbb{E}_2 \log \cosh g_0(\resub{\m,\,}\q_1, \q_0) \tanh g_0(\resub{\m,\,}\q_1, \q_0)\right] \notag \\
&- \mathbb{E}_1 \left[ \mathbb{E}_2 \log \cosh g_0(\resub{\m,\,}\q_1, \q_0) \left( \mathbb{E}_2 \tanh g_0(\resub{\m,\,}\q_1, \q_0)\right)^2\right] \notag \\
&+ \dfrac{2\beta^2 \alpha \q_0(\q_1-\q_0)}{\Delta_1(\q_1)^3} \mathbb{E}_1 \left[ \mathbb{E}_2 \left( 1-\tanh^2 g_0(\resub{\m,\,}\q_1, \q_0)\right)\right] \notag \\
&- \dfrac{4\beta^3 \alpha \q_0 (\q_1-\q_0)}{\Delta_1(\q_1)^3} \mathbb{E}_1 \left[\mathbb{E}_2 \tanh g_0(\resub{\m,\,}\q_1, \q_0)(1-\tanh^2 g_0(\resub{\m,\,}\q_1, \q_0)) \mathbb{E}_2 \tanh^2 g_0(\q_1, \q_0) \right] \notag \\
&+ \dfrac{2\beta^3 \alpha \q_0 (\q_1-\q_0)}{\Delta_1(\q_1)^3} \mathbb{E}_1 \left[ \mathbb{E}_2 \tanh^2 g_0(\resub{\m,\,}\q_1, \q_0) \mathbb{E}_2 \left( 1-\tanh^2 g_0(\resub{\m,\,}\q_1, \q_0)\right)\right] \notag \\
&+ \dfrac{\beta^2 \alpha (\q_1-\q_0)^2}{\Delta_1(\q_1)^3} \mathbb{E}_1 \left[ \mathbb{E}_2 \tanh g_0(\resub{\m,\,}\q_1, \q_0) \mathbb{E}_2 \tanh g_0(\resub{\m,\,}\q_1, \q_0)(1-\tanh^2 g_0(\resub{\m,\,}\q_1, \q_0))\right]\Big\}
% \mathbb{E}_1 \left[ \mathbb{E}_2 \tanh g_0(\q_0, \q) \mathbb{E}_2 \tanh g_0(\q_0, \q) \ln \cosh g_0(\q_0, \q)\right] \notag \\
%     &- \mathbb{E}_1 \left[ \left(\mathbb{E}_2 \tanh g_0(\q_0, \q)\right)^2 \mathbb{E}_2 \ln \cosh g_0(\q_0, \q) \right] \notag \\
%     &
%     %\textcolor{white}{\mathbb{E}_1\left( \mathbb{E}_2 \tanh g \right)^2 + 2 \theta}
%     + \dfrac{\beta^3 (\q-\q_0)\alpha \q_0 }{\Delta_1(\qq, \q)^3} \mathbb{E}_1 \left[ \mathbb{E}_2 (1-\tanh^2 g_0(\q_0, \q)) \mathbb{E}_2 \tanh^2 g_0(\q_0, \q) \right] \notag \\
%     &
%     %\textcolor{white}{\mathbb{E}_1\left( \mathbb{E}_2 \tanh g \right)^2 + 2 \theta}
%     + \dfrac{2\beta^3 \alpha \q (\q\!-\!\q_0) }{\Delta_1(\q, \qq)^3} \mathbb{E}_1 \left[ \mathbb{E}_2 \tanh g \mathbb{E}_2 \tanh g_0(\q_0, \q) (1\!-\!\tanh^2 g_0(\q_0, \q)) \right]\!\!\Big\}
    \label{eq:B-Hopfield}
\end{align}
and 
\begin{align}
C(\resub{\m,\,} \q,\q_0) =& \dfrac{\alpha \beta^3 \q_1 (\q_1-\q_0)}{(1-\beta(1-\q_1))^3} \mathbb{E} \left[ \left( 1- \tanh^2 g_0(\resub{\m,\,}\q_1, \q_0) \right) \log \cosh g_0(\resub{\m,\,}\q_1, \q_0) \right.\notag \\
&\left.+ \tanh^2 g_0(\resub{\m,\,}\q_1, \q_0) - 2 \tanh g_0(\resub{\m,\,}\q_1, \q_0) + 2 \tanh^3 g_0(\resub{\m,\,}\q_1, \q_0)\right]
    \label{eq:C-Hopfield}
\end{align}
where $g_0(\resub{\m,\,}\q_0, \q_1)$ is defined in \eqref{eq:g0HOP}.
\item 
For Hebbian networks with $P$-node interactions,
the subleading contributions to the overlaps $\q_0$ and $\q_1$ in the expansion around $\theta=1$ are given by 
\begin{align}
A(\resub{\m,\,}\q_0,\q_1)=& 2\mathbb{E}_1\left\{\dfrac{\mathbb{E}_2 \ln \cosh g(\resub{\m,\,}\q_1, \q_0) \sinh g(\resub{\m,\,}\q_1, \q_0) \tanh g(\resub{\m,\,}\q_1, \q_0)}{\mathbb{E}_2 \cosh g(\resub{\m,\,}\q_1, \q_0)  }\right\}- \notag \\
&2\mathbb{E}_1 \left\{\dfrac{\mathbb{E}_2 \ln \cosh g(\resub{\m,\,}\q_1, \q_0) \cosh g(\resub{\m,\,}\q_1, \q_0) \mathbb{E}_2 \sinh g(\resub{\m,\,}\q_1, \q_0) \tanh g(\resub{\m,\,}\q_1, \q_0)}{\left(\mathbb{E}_2 \cosh g(\resub{\m,\,}\q_1, \q_0) \right)^2 }\right\},
    \label{eq:A-Dense_t1}
    \end{align}
    \begin{align}
B(\resub{\m,\,}\q_0,\q_1)=& 2\mathbb{E}_1 \left\{ \dfrac{\mathbb{E}_2 \sinh g(\resub{\m,\,}\q_1, \q_0) \tanh  g(\resub{\m,\,}\q_1, \q_0) \mathbb{E}_2 \cosh g(\resub{\m,\,}\q_1, \q_0) \log \cosh  g(\resub{\m,\,}\q_1, \q_0) }{\left( \mathbb{E}_2 \cosh  g(\resub{\m,\,}\q_1, \q_0) \right)^2} \right\}\notag \\
&-2\mathbb{E}_1 \left\{ \dfrac{\left(\mathbb{E}_2 \sinh g(\resub{\m,\,}\q_1, \q_0) \tanh  g(\resub{\m,\,}\q_1, \q_0) \right)^2\mathbb{E}_2 \sinh g(\resub{\m,\,}\q_1, \q_0) \log \cosh  g(\resub{\m,\,}\q_1, \q_0) }{\left( \mathbb{E}_2 \cosh  g(\resub{\m,\,}\q_1, \q_0) \right)^3} \right\},
    \label{eq:B-Dense_t1}
    \end{align}
    \begin{align}
        C(\resub{\m,\,}\q_0,\q_1)=& \mathbb{E}_1 \left[ \dfrac{\mathbb{E}_2 \sinh g(\resub{\m,\,}\q_1, \q_0)\log \cosh g(\resub{\m,\,}\q_1, \q_0)}{\mathbb{E}_2 \cosh g(\resub{\m,\,}\q_1, \q_0) }\right] - \notag \\
        & \mathbb{E}_1 \left[ \dfrac{\mathbb{E}_2 \sinh g(\resub{\m,\,}\q_1, \q_0)\mathbb{E}_2 \cosh g(\resub{\m,\,}\q_1, \q_0) \log \cosh g(\resub{\m,\,}\q_1, \q_0)}{\left(\mathbb{E}_2 \cosh g(\resub{\m,\,}\q_1, \q_0)\right)^2 }\right]
         \label{eq:C-Dense_t1}
    \end{align}
    respectively, 
whereas, for the expansion around $\theta=0$ they evaluate to 
\begin{align}
A(\resub{\m,\,}\q_0,\q_1)=& \mathbb{E}_1\mathbb{E}_2 \ln \cosh g(\resub{\m,\,}\q_1, \q_0) \tanh^2 g(\resub{\m,\,}\q_1, \q_0)\notag \\
&- \mathbb{E}_1 \left( \mathbb{E}_2 \ln \cosh g(\resub{\m,\,}\q_1, \q_0)  \mathbb{E}_2 \tanh^2 g(\resub{\m,\,}\q_1, \q_0)  \right)
    \label{eq:A-Dense}
    \end{align}
    \begin{align}
B(\resub{\m,\,}\q_0,\q_1)=& 
\left\{ 2 \mathbb{E}_1 \left[ \mathbb{E}_2 \tanh g(\resub{\m,\,}\q_1, \q_0) \mathbb{E}_2 \ln \cosh g(\resub{\m,\,}\q_1, \q_0) \tanh g(\resub{\m,\,}\q_1, \q_0) \right] \right.\notag \\
    &\left.- 2 \mathbb{E}_1 \left[ \left( \mathbb{E}_2 \tanh g(\resub{\m,\,}\q_1, \q_0) \right)^2 \mathbb{E}_2 \ln \cosh g(\resub{\m,\,}\q_1, \q_0) \right]\right\},
    \label{eq:B-Dense}
    \end{align}
    \begin{align}
        C(\resub{\m,\,}\q_0,\q_1)=& \mathbb{E}_1 \left[ \mathbb{E}_2 \tanh g(\resub{\m,\,}\q_1, \q_0) \log \cosh g(\resub{\m,\,}\q_1, \q_0) \right] \notag \\
        &-\mathbb{E}_1 \left[ \mathbb{E}_2 \tanh g(\resub{\m,\,}\q_1, \q_0) \mathbb{E}_2 \log \cosh g(\resub{\m,\,}\q_1, \q_0) \right]
        \label{eq:C-Dense}
    \end{align}
where $g(\resub{\m,\,}\q_1, \q_0)$ is defined in \eqref{eq:gDense}.
\item For the Sherrington-Kirkpatrick model the 
subleading contributions to the overlaps $\q_0$ and $\q_1$ in the expansion around $\theta=1$ are given by 
\begin{align}
    A(\q_0, \q_1)= \mathbb{E}_1 &\left\{ \dfrac{\mathbb{E}_2 \log \cosh g(\q_0, \q_1) \sinh g(\q_0, \q_1) \tanh g(\q_0, \q_1) \mathbb{E}_2 \cosh g(\q_0, \q_1) }{(\mathbb{E}_2 \cosh g )^2}\right. \notag \\
    &\left. -\dfrac{\mathbb{E}_2 \sinh g(\q_0, \q_1) \tanh g(\q_0, \q_1) \mathbb{E}_2 \cosh g(\q_0, \q_1) \log \cosh g(\q_0, \q_1)}{(\mathbb{E}_2 \cosh g(\q_0, \q_1) )^2}\right\} \label{eq:A_Pspin_t1_SK}\\
    B(\q_0, \q_1) =& \mathbb{E}_1 \left\{ \dfrac{\mathbb{E}_2 \log \cosh g(\q_0, \q_1) \sinh g(\q_0, \q_1) \mathbb{E}_2 \cosh g(\q_0, \q_1) }{(\mathbb{E}_2 \cosh g(\q_0, \q_1) )^2}\right. \notag \\
    &\left.- \dfrac{ \mathbb{E}_2 \sinh g(\q_0, \q_1) \mathbb{E}_2 \cosh g(\q_0, \q_1) \log \cosh g(\q_0, \q_1)}{(\mathbb{E}_2 \cosh g(\q_0, \q_1) )^2}\right\} \label{eq:B_Pspin_t1_SK}.
\end{align}
where $g(\q_1, \q_0)$ is defined in \eqref{eq:gSK}.
\iffalse
whereas, in the expansion around $\theta=0$ they evaluate to 
\begin{align}
\label{eq:A-SK}
     A(\q_1, \q_0) =&\mathbb{E}_1 \left[ \mathbb{E}_2 \ln \cosh g(\q_1, \q_0) \tanh^2 g(\q_1, \q_0) \right] \notag \\
     &- \mathbb{E}_1 \left[ \mathbb{E}_2 \ln \cosh g(\q_1, \q_0) \mathbb{E}_2 \tanh^2 g(\q_1, \q_0) \right]\\
    \label{eq:B-SK}
    \end{align}
    and 
    \begin{align}
    B( \q_1, \q_0)=& 2\left\{ \mathbb{E}_1 \left( \mathbb{E}_2 \tanh g(\q_1, \q_0) \mathbb{E}_2 \ln \cosh g(\q_1, \q_0) \tanh g(\q_1, \q_0)\right) \right. \notag \\
    &\left. 
    - \mathbb{E}_1 \left( \left(\mathbb{E}_2 \tanh g(\q_1, \q_0) \right)^2\mathbb{E}_2 \ln \cosh g(\q_1, \q_0) \right)\right\}.
\end{align}
where $g(\q_1, \q_0)$ is defined as \eqref{eq:gSK}.
\fi
\item For the Ising P-spin model, these terms, in the expansion around $\theta=1$, evaluate to
\begin{align}
    A(\q_0, \q_1)=& \mathbb{E}_1 \left\{ \dfrac{\mathbb{E}_2 \log \cosh g(\q_1, \q_0) \sinh g(\q_1, \q_0) \tanh g(\q_1, \q_0) \mathbb{E}_2 \cosh g(\q_1, \q_0)}{(\mathbb{E}_2 \cosh g(\q_1, \q_0) )^2}\right. \notag \\
    &\left.-\dfrac{\mathbb{E}_2 \sinh g(\q_1, \q_0) \tanh g(\q_1, \q_0) \mathbb{E}_2 \cosh g(\q_1, \q_0) \log \cosh g(\q_1, \q_0)}{(\mathbb{E}_2 \cosh g(\q_1, \q_0) )^2}\right\} \label{eq:A_Pspin_t1}\\
    B(\q_0, \q_1) =& \mathbb{E}_1 \left\{ \dfrac{\mathbb{E}_2 \log \cosh g(\q_1, \q_0) \sinh g(\q_1, \q_0) \mathbb{E}_2 \cosh g(\q_1, \q_0) }{(\mathbb{E}_2 \cosh g(\q_1, \q_0) )^2}\right. \notag \\
    &\left.-\dfrac{\mathbb{E}_2 \sinh g(\q_1, \q_0) \mathbb{E}_2 \cosh g(\q_1, \q_0) \log \cosh g(\q_1, \q_0)}{(\mathbb{E}_2 \cosh g(\q_1, \q_0) )^2}\right\} \label{eq:B_Pspin_t1}.
\end{align}
where $g(\q_1, \q_0)$ is defined in \eqref{eq:gPSK}.
\item Finally, for the spherical $P$-spin model, the contributions to linear orders in the expansion around $\theta=1$ (see \eqref{eq:SPSq1-A} and 
\eqref{eq:q1SGS}) are
    \begin{align}
    \label{eq:A-SPS}
     A(\q_0, \q_1) =&\dfrac{2\q_0(\q_0-\q_1)}{(1-\q_0)^3} \\
      B(\q_0, \q_1)=&  \dfrac{(\q_1-\q_0)^2}{(1-\q_0)(1-\q_1)} +\dfrac{2\q_0(\q_0-\q_1)}{(1-\q_0)^3}.
     \label{eq:B-SPS}
\end{align}
while the contributions to linear orders in the expansion around $\theta=0$ (see \eqref{eq:SPS-theta0-1} and 
\eqref{eq:SPS-theta0-0}) are: 
\begin{align}
\label{eq:A-SPS0}
A(\q_0, \q_1)=& \dfrac{(\q_1^2-\q_0^2)}{(1-\q_1)^3}\\
B(\q_0, \q_1)=&\dfrac{2 \q_0(\q_1-\q_0)}{(1-\q_1)^3} \label{eq:B-SPS0}
\end{align}
    \end{itemize}

\begin{thebibliography}{10}

\bibitem{AABO-JPA2020}
E.~Agliari, L.~Albanese, A.~Barra, and G.~Ottaviani.
\newblock Replica symmetry breaking in neural networks: A few steps toward
  rigorous results.
\newblock {\em Journal of Physics A: Mathematical and Theoretical}, 53, 2020.

\bibitem{Barra-PRLdetective}
E.~Agliari, F.~Alemanno, A.~Barra, M.~Centonze, and A.~Fachechi.
\newblock Neural networks with a redundant representation: Detecting the
  undetectable.
\newblock {\em Physical Review Letters}, 124:28301, 2020.

\bibitem{Fachechi1}
E.~Agliari, F.~Alemanno, A.~Barra, and A.~Fachechi.
\newblock Generalized guerra's interpolation schemes for dense associative
  neural networks.
\newblock {\em Neural Networks}, 128:254--267, 2020.

\bibitem{Albanese2021}
L.~Albanese, F.~Alemanno, A.~Alessandrelli, and A.~Barra.
\newblock Replica symmetry breaking in dense hebbian neural networks.
\newblock {\em Journal of Statistical Physics}, 189(2):1--41, 2022.

\bibitem{Amit}
D.~J. Amit.
\newblock {\em Modeling brain function: The world of attractor neural
  networks}.
\newblock Cambridge university press, 1989.

\bibitem{AnnibaleGualdiCavagna04}
A.~Annibale, G.~Gualdi, and A.~Cavagna.
\newblock Coexistence of supersymmetric and supersymmetry-breaking states in
  spherical spin-glasses.
\newblock {\em J. Phys. A: Math. Gen. 37 11311}, 37(47):11311, 2004.

\bibitem{Antenucci15b}
F.~Antenucci, C.~Conti, A.~Crisanti, and L.~Leuzzi.
\newblock General phase diagram of multimodal ordered and disordered lasers in
  closed and open cavities.
\newblock {\em Phys. Rev. Lett.}, 114:043901, Jan 2015.

\bibitem{Antenucci16}
F.~Antenucci, A.~Crisanti, M.~Ibáñez-Berganza, A.~Marruzzo, and L.~Leuzzi.
\newblock Statistical mechanics models for multimode lasers and random lasers.
\newblock {\em Philosophical Magazine}, 96(7-9):704--731, 2016.

\bibitem{Antenucci15}
F.~Antenucci, A.~Crisanti, and L.~Leuzzi.
\newblock Complex spherical 2 + 4 spin glass: A model for nonlinear optics in
  random media.
\newblock {\em Phys. Rev. A}, 91:053816, May 2015.

\bibitem{Antenucci15c}
F.~Antenucci, M.~Ib\'a\~nez Berganza, and L.~Leuzzi.
\newblock Statistical physics of nonlinear wave interaction.
\newblock {\em Phys. Rev. B}, 92:014204, Jul 2015.

\bibitem{Antenucci21}
F.~Antenucci, G.~Lerario, B.~S. Fernand\'ez, L.~De~Marco, M.~De~Giorgi,
  D.~Ballarini, D.~Sanvitto, and L.~Leuzzi.
\newblock Demonstration of self-starting nonlinear mode locking in random
  lasers.
\newblock {\em Phys. Rev. Lett.}, 126:173901, Apr 2021.

\bibitem{Zecchina-Pnas2016}
C.~Baldassi, C.~Borgs, J.~T. Chayes, A.~Ingrosso, C.~Lucibello, L.~Saglietti,
  and R.~Zecchina.
\newblock Unreasonable effectiveness of learning neural networks: From
  accessible states and robust ensembles to basic algorithmic schemes.
\newblock {\em Proceedings of the National Academy of Sciences},
  113(48):E7655--E7662, 2016.

\bibitem{Zecchina-PRL2021}
C.~Baldassi, C.~Lauditi, E.~M. Malatesta, G.~Perugini, and R.~Zecchina.
\newblock Unveiling the structure of wide flat minima in neural networks.
\newblock {\em Physical Review Letters}, 127(27):278301, 2021.

\bibitem{bardina2004}
X.~Bardina, D.~M\'arquez-Carreras, C.~Rovira, and S.~Tindel.
\newblock The p-spin interaction model with external field.
\newblock {\em Potential Analysis}, 21:311–362, 2004.

\bibitem{glassy}
A.~Barra, G.~Genovese, F.~Guerra, and D.~Tantari.
\newblock How glassy are neural networks?
\newblock {\em Journal of Statistical Mechanics: Theory and Experiment},
  2012(07):P07009, 2012.

\bibitem{Cavagna}
T.~Castellani and A.~Cavagna.
\newblock Spin-glass theory for pedestrians.
\newblock {\em Journal of Statistical Mechanics: Theory and Experiment},
  2005(05):P05012, 2005.

\bibitem{Charbonneau-PRE2019}
P.~Charbonneau, Y.~Hu, A.~Raju, J.~P. Sethna, and S.~Yaida.
\newblock Morphology of renormalization-group flow for the de
  almeida--thouless--gardner universality class.
\newblock {\em Physical Review E}, 99(2):022132, 2019.

\bibitem{chen2021almeida}
W.-K. Chen.
\newblock On the almeida-thouless transition line in the
  sherrington-kirkpatrick model with centered gaussian external field.
\newblock {\em Electronic Communications in Probability}, 26:1--9, 2021.

\bibitem{coolen2001statistical}
A.~Coolen.
\newblock Statistical mechanics of recurrent neural networks i—statics.
\newblock In {\em Handbook of biological physics}, volume~4, pages 553--618.
  Elsevier, 2001.

\bibitem{Coolen}
A.~C.~C. Coolen, R.~Kühn, and P.~Sollich.
\newblock {\em Theory of neural information processing systems}.
\newblock OUP Oxford, 2005.

\bibitem{Crisanti-RSB}
A.~Crisanti, D.~J. Amit, and H.~Gutfreund.
\newblock Saturation level of the hopfield model for neural network.
\newblock {\em Europhysics Letters (EPL)}, 2:337--341, 8 1986.

\bibitem{CrisantiLeuzzi04}
A.~Crisanti and L.~Leuzzi.
\newblock Spherical $2+p$ spin-glass model: An exactly solvable model for glass
  to spin-glass transition.
\newblock {\em Phys. Rev. Lett.}, 93:217203, Nov 2004.

\bibitem{CrisantiLeuzzi05}
A.~Crisanti and L.~Leuzzi.
\newblock Spherical $2+p$ spin-glass model: An analytically solvable model with
  a glass-to-glass transition.
\newblock {\em Phys. Rev. B}, 73:014412, Jan 2006.

\bibitem{Crisanti}
A.~Crisanti and H.~J. Sommers.
\newblock The spherical p-spin interaction spin glass model: the statics.
\newblock {\em Zeitschrift für Physik B Condensed Matter}, 87:341--354, 1992.

\bibitem{de1978stability}
J.~R. de~Almeida and D.~J. Thouless.
\newblock Stability of the sherrington-kirkpatrick solution of a spin glass
  model.
\newblock {\em Journal of Physics A: Mathematical and General}, 11(5):983,
  1978.

\bibitem{FolenaFranzTersenghi20}
G.~Folena, S.~Franz, and F.~Ricci-Tersenghi.
\newblock Rethinking mean-field glassy dynamics and its relation with the
  energy landscape: The surprising case of the spherical mixed p-spin model.
\newblock {\em Phys. Rev. X}, 10:031045, Aug 2020.

\bibitem{gardner1985spin}
E.~Gardner.
\newblock Spin glasses with p-spin interactions.
\newblock {\em Nuclear Physics B}, 257:747--765, 1985.

\bibitem{Gardner}
E.~Gardner.
\newblock Multiconnected neural network models.
\newblock {\em Journal of Physics A: General Physics}, 20, 1987.

\bibitem{guerra_broken}
F.~Guerra.
\newblock Broken replica symmetry bounds in the mean field spin glass model.
\newblock {\em Communications in Mathematical Physics}, 233:1--12, 2003.

\bibitem{guerra2006replica}
F.~Guerra.
\newblock The replica symmetric region in the sherrington-kirkpatrick mean
  field spin glass model. the almeida-thouless line.
\newblock {\em arXiv preprint cond-mat/0604674}, 2006.

\bibitem{Gavin-PRE2022}
G.~S. Hartnett, E.~Parker, and E.~Geist.
\newblock Replica symmetry breaking in bipartite spin glasses and neural
  networks.
\newblock {\em Physical Review E}, 98(2):022116, 2018.

\bibitem{Holler-PRE2020}
J.~H{\"o}ller and N.~Read.
\newblock One-step replica-symmetry-breaking phase below the de
  almeida--thouless line in low-dimensional spin glasses.
\newblock {\em Physical Review E}, 101(4):042114, 2020.

\bibitem{HopKro1}
D.~Krotov and J.~J. Hopfield.
\newblock Dense associative memory for pattern recognition.
\newblock {\em Advances in Neural Information Processing Systems}, pages
  1180--1188, 2016.

\bibitem{Chokri-JSP2022}
C.~Manai and S.~Warzel.
\newblock The de almeida--thouless line in hierarchical quantum spin glasses.
\newblock {\em Journal of Statistical Physics}, 186(1):1--32, 2022.

\bibitem{nishimori2001statistical}
H.~Nishimori.
\newblock {\em Statistical physics of spin glasses and information processing:
  an introduction}.
\newblock Number 111. Clarendon Press, 2001.

\bibitem{sherrington1975solvable}
D.~Sherrington and S.~Kirkpatrick.
\newblock Solvable model of a spin-glass.
\newblock {\em Physical review letters}, 35(26):1792, 1975.

\bibitem{Steffan-RSB}
H.~Steffan and R.~Kühn.
\newblock Replica symmetry breaking in attractor neural network models.
\newblock {\em Zeitschrift für Physik B Condensed Matter}, 95, 1994.

\bibitem{talagrand2003spin}
M.~Talagrand et~al.
\newblock {\em Spin glasses: a challenge for mathematicians: cavity and mean
  field models}, volume~46.
\newblock Springer Science \& Business Media, 2003.

\bibitem{Kondor-arxiv2022}
T.~Temesv{\'a}ri and I.~Kondor.
\newblock Field theory for the almeida-thouless transition.
\newblock {\em arXiv preprint arXiv:2212.01654}, 2022.

\bibitem{Thouless86}
D.~J. Thouless.
\newblock Spin-glass on a bethe lattice.
\newblock {\em Phys. Rev. Lett.}, 56:1082--1085, Mar 1986.

\bibitem{toninelli2002almeida}
F.~L. Toninelli.
\newblock About the almeida-thouless transition line in the
  sherrington-kirkpatrick mean-field spin glass model.
\newblock {\em EPL (Europhysics Letters)}, 60(5):764, 2002.

\bibitem{TonoloNieddaGradenigo}
T.~Tonolo, J.~Niedda, and G.~Gradenigo.
\newblock Marginal stability in the spherical spin glass: on the competition
  between disorder and non-linearity.
\newblock {\em In preparation}.

\bibitem{Haiping-PRR2023}
Y.~Zhao, J.~Qiu, M.~Xie, and H.~Huang.
\newblock Equivalence between belief propagation instability and transition to
  replica symmetry breaking in perceptron learning systems.
\newblock {\em Physical Review Research}, 4(2):023023, 2022.

\end{thebibliography}
\end{document}